\documentclass[12pt]{iopart}
\usepackage{iopams}  
\usepackage{xcolor}
\usepackage[dvips]{graphicx}
\usepackage{wrapfig}
\newcommand{\sbt}{\,\begin{picture}(-1,1)(-1,-3)\circle*{2.2}\end{picture}\ }
\newcommand{\iint}{\int \!\!\! \int}

\newlength{\Width}
\newlength{\Height}
\newlength{\Depth}
\usepackage[hang,small,bf]{caption}
\usepackage[subrefformat=parens]{subcaption}
\begin{document}
\hfill{NIT-KMP-202201}

\title{Nonlinear $O(3)$ sigma model in discrete complex analysis}

\author{M{\scriptsize ASARU} K{\scriptsize AMATA}$\dagger$, 
M{\scriptsize ASAYOSHI} S{\scriptsize EKIGUCHI}$\ddagger$ {\scriptsize AND} Y{\scriptsize UUKI} T{\scriptsize ADOKORO}$\S$}

\address{\it Department of Mathematics, National Institute of Technology, Kisarazu College,
2-11-1 Kiyomidai-Higashi, Kisarazu, Chiba, 292-0041, Japan}

\ead{$^{\dagger}$kamata@kisarazu.ac.jp, $^{\ddagger}$masa@kisarazu.ac.jp, $^{\S}$tado@nebula.n.kisarazu.ac.jp}
\vspace{10pt}
\begin{indented}
\item[] 
\end{indented}
\color{black}

\begin{abstract}
We  present a discrete version of the two-dimensional nonlinear $O(3)$ sigma 
model examined by Belavin and Polyakov. We formulate it by means of Mercat's 
discrete complex analysis and its elaboration by Bobenko and G{\"u}nther. 
We define a weighted discrete Dirichlet energy and area on a planar quad-graph 
and derive an inequality between them. 
We write $f$ for the complex function obtained from the unit vector field of the model.  
The inequality is saturated if and only if the $f$ is discrete (anti-)holomorphic.
By using a weight $W$ obtained from a kind of tiling of the sphere $S^2$, the weighted 
discrete area ${\cal A}^{W}_{\diamondsuit}(f)$ admits a geometrical interpretation, 
namely, ${\cal A}^{W}_{\diamondsuit}(f)=4 \pi N $ for a topological quantum number $N \in \pi_2(S^2)$. 
This ensures the topological stability of the solution described by the $f$, and we have 
the quantized energy $E^{W}_{\diamondsuit}(f)=|{\cal A}^{W}_{\diamondsuit}(f)|=4 \pi |N| $.
For quad-graphs with orthogonal diagonals, we show that the discrete (anti-)holomorphic 
function $f$ satisfies the Euler--Lagrange equation derived from the weighted discrete 
Dirichlet energy. On some rhombic lattices, the discrete power functions $z^{(N)}$ give 
the topological quantum number $N$. 
Moreover, the weighted discrete Dirichlet energy, area, and Euler--Lagrange equation 
tend to their continuous forms as the lattice spacings tend to zero. 
\end{abstract}


\section{Introduction}
There exist prominent topological objects in a certain class of field theories.
In four-dimensional Euclidean space ${\mathbb{R}^4}$, 
the Belavin-Polyakov-Schwarz-Tyupkin (BPST) instanton \cite{BPST} is a solution to 
the (anti-)self-dual Yang--Mills equations $F_{\mu\nu}=\pm \,{*}F_{\mu\nu}$ 
that minimizes the action 
\begin{equation}
S=\frac{1}{2}\int_{\mathbb{R}^4}\! d^4x\, {\tr}(F_{\mu\nu}F^{\mu\nu})
={\|}F\mp \,{*}F{\|}^2\pm\,8\pi^2 N\ge 8\pi^2|N|, 
\label{eq_ineqS_4d}
\end{equation}
where ${\|}\cdot{\|}$ is some norm and $N \in \pi_3(S^3)$ is the instanton number. Similarly, 
the Bogomol'nyi-Prasad-Sommerfield (BPS) monopole \cite{Bogo} in ${\mathbb{R}^3}$ is 
a solution to the (anti-)Bogomol'nyi equations $F_{jk}=\pm\varepsilon_{jkl}D_{l}\phi $ that minimizes 
the energy 
\begin{equation}
\fl \hspace{1.2cm} E=\frac{1}{2}\int_{\mathbb{R}^3}\! d^3x\,{\tr}\left[{\mbox{\boldmath $B$}}^2+
({\mbox{\boldmath $D$}}\phi)^2\right] 
={\|}F\mp \,{*}D\phi {\|}^2 \pm 4\pi |\langle\phi\rangle| N\ge 4\pi |\langle\phi\rangle||N|, 
\label{eq_ineqS_3d}
\end{equation}
where $\mbox{\boldmath $B$}$ is the magnetic field, $\langle\phi\rangle$ is the vacuum 
expectation value of the Higgs scalar $\phi$, and $N \in \pi_2(S^2)$ is a topological quantum number 
proportional to the magnetic charge. 

In ${\mathbb{R}^2}$, the nonlinear $O(3)$ sigma model (or the ${\mathbb{C}}P^{\mathcal{N}-1}$ 
model with $\mathcal{N}=2$) has properties analogous to those of instantons and monopoles. 
Indeed, we can show that the energy $E$ of the model fulfills the following inequality \cite{BelavPolyak} 
\begin{equation}
E=\frac{1}{2}\int_{\mathbb{R}^2}\! d^2x\, \partial_{\mu}\mbox{\boldmath $n$}\sbt\, 
\partial^{\mu}\mbox{\boldmath $n$}
= \Biggl{\|}\frac{df \mp i*df}{1+|f|^2}\Biggr{\|}^2 \pm\,4\pi N\ge 4\pi |N|, 
\label{eq_E_ineq} 
\end{equation}
where $\mbox{\boldmath $n$}:=(n^1, n^2, n^3)$ is a three-dimensional vector with 
the constraint $\mbox{\boldmath $n$}\sbt\, \mbox{\boldmath $n$}=1$, $\mu=1, 2$, and ${\|}\cdot{\|}$ is 
the standard norm. A function $f$ of $z=x+iy \in \mathbb{C}$ is defined as 
the composition $f:={\rm St} \circ {\mbox{\boldmath$n$}}$ of the unit vector $\mbox{\boldmath $n$}$ 
and the stereographic projection ${\rm St}$ 
from a unit sphere $S^2$ to $\mathbb{C}$. The explicit form of $f$ is given by \cite{BelavPolyak,Raja,MantonSut}
\begin{equation}
\fl \hspace{1.5cm} f: \mathbb{R}^2 \stackrel{\mbox{\boldmath $n$}}{\longrightarrow}  
S^2  \stackrel{\rm St}{\longrightarrow}
\mathbb{R}^2 \cong \mathbb{C}, \,\,\,\,(x,y) \mapsto f(x,y)=\frac{n^1(x,y)+in^2(x,y)}{1+n^3(x,y)}. 
\label{eq_f(x,y)}
\end{equation}
Appropriate boundary conditions must be applied to the vector $\mbox{\boldmath $n$}$ at infinity 
to ensure that the energy $E$ remains finite. 
The topological quantum number $N \in \pi_2(S^2)$, which classifies the map 
\mbox{\boldmath $n$\,:} $S^2 \cong \mathbb{C}\cup\{\infty\} \to S^2$, is defined as
\begin{equation}
N:=\frac{1}{4\pi}\int_{\mathbb{R}^2}\! d^2x\,\, {\mbox{\boldmath $n$}}\sbt (\partial_{x} {\mbox{\boldmath $n$}}\times \partial_{y} {\mbox{\boldmath $n$}})
=\frac{1}{4\pi}\int_{\mathbb{R}^2} {\mbox{\boldmath $n$}}^{*}d{\rm vol}_{S^2}. 
\label{eq_N} 
\end{equation}
If the complex function $f$ is (anti-)holomorphic, then the Cauchy--Riemann equations 
(or their anti-versions) hold; equivalently, in a coordinate-independent manner, $df \mp i*df=0$ holds. 
The inequality (\ref{eq_E_ineq}) is saturated if and only if the function $f$ is (anti-)holomorphic. 
At saturation, the equality $E=4\pi |N|$ is attained, and we have an (anti-)instanton solution. 
A general solution for instantons with arbitrary $N \in \mathbb{N}$ is obtained in \cite{BelavPolyak}.

If we express the energy $E$ and the topological quantum number $N$ in terms of $f$, we have 
\begin{eqnarray}
E&=& 2\int_{\mathbb{R}^2}\! d^2x 
\frac{|\partial_{x} f|^2+|\partial_{y} f|^2}{(1+|f|^2)^2}, \label{eq_cont_E} \\
N&=&\frac{1}{\pi}\int_{\mathbb{R}^2} d^2x\,\, 
\frac{{\rm Im}(\partial_{y} f\cdot \partial_{x}\bar{f})}{(1+|f|^2)^2}=\frac{1}{4\pi}\cal{A}, \label{eq_cont_A} 
\end{eqnarray}
where ${\cal{A}}$ is the spherical area swept out by the vector $\mbox{\boldmath $n$}$. 
The energy $E$ in (\ref{eq_cont_E}), which is unconstrained, gives the continuous Euler--Lagrange equation 
\begin{eqnarray}
\fl \hspace{2cm} {[{\rm EL}]}^{\rm cont.}(z):&=&-\frac{\partial E}{\partial \bar{f}(z)} \nonumber \\
&=&\frac{8}{(1+|f|^2)^2}\Biggl(\partial_{z}\partial_{\bar{z}}f-\frac{2\bar{f}}{1+|f|^2}\partial_{z}f \cdot \partial_{\bar{z}}f\Biggr)=0. 
\label{eq_ELeqf_BG_conti}
\end{eqnarray}
Here, $4\partial_{z}\partial_{\bar{z}}=\Delta$ is the two-dimensional Laplacian. The expression (\ref{eq_ELeqf_BG_conti}) 
shows that holomorphic 
($\partial_{\bar{z}}f=0$) or anti-holomorphic ($\partial_{z}f=0$) functions can both satisfy the continuous 
Euler--Lagrange equation ${[{\rm EL}]}^{\rm cont.}(z)=0$. 

The construction of a discrete model on a lattice such that it ensures the topological stability of 
solutions is a challenging problem that has been the object of study for decades.
Berg and L{\"u}scher \cite{BerLu81} defined a topological quantum number in the lattice $O(3)$ sigma model
by using the Gauss--Bonnet formula. 
Speight and Ward \cite{SpeightWard94} presented a discrete sine-Gordon system with a Bogomol'nyi lower 
bound on the energy. 
Piette, Schroers, and Zakrzewski \cite{PietteSchrZakrz95}, and Ward \cite{Ward95} 
studied a two-dimensional stable topological skyrmions with the target space $S^2$. 
Leese \cite{Leese89} derived a discrete Bogomol'nyi equation for  
the nonlinear $O(3)$ sigma model in 2+1 dimensions where the energy density is 
radially symmetric. Ward \cite{Ward97} also studied a lattice abelian Higgs system. 

On the other hand, discrete complex analysis has a somewhat long history. There exist many concepts and 
theorems corresponding to those in ordinary complex analysis, such as discrete holomorphic functions, 
harmonic functions, and the Cauchy integral theorem. Mercat \cite{Mer01,Mer07} used arbitrary two-dimensional 
metric graphs and their duals, and Wilson \cite{Wilson} used a triangular cellular decomposition of the Riemann surface. 
Each obtained discrete period matrices, which tend to the usual period matrices in a continuous limit. 
For the convergence proof of discrete period matrices, see \cite{BobeSkop}. 
Bobenko and Suris \cite{BobeSur08} developed a discrete differential geometry and applied it to 
the discrete integrable systems. Bobenko and G{\"u}nther \cite{BobeGunt16, BobeGunt17} 
presented a theory of discrete Riemann surfaces in which the 
medial graph on a quad-graph played a central role. A more detailed description of this history is 
presented in the Introduction of \cite{BobeGunt16}. 

The aim of this paper is to construct a discrete version of the 
nonlinear $O(3)$ sigma model on two-dimensional lattices where the topological stability of the solutions is ensured. 
We use the discrete complex analysis of Mercat \cite{Mer01,Mer07} and its elaboration by 
Bobenko and G{\"u}nther \cite{BobeGunt16} based on the medial graph of 
planar quad-graphs. They defined a discrete Dirichlet energy $E_{\diamondsuit}(f):=\langle df, df \rangle$ and 
derived a discrete Laplace equation as the Euler--Lagrange equation. 
In this paper, we introduce a weight function $W$ and define a weighted discrete Dirichlet 
energy $E^{W}_{\diamondsuit}(f):=2\langle Wdf, df \rangle$ and a weighted discrete 
area ${\cal A}^{W}_{\diamondsuit}(f):=-2i\langle Wdf, \star df \rangle$. We derive an inequality 
between them, which is saturated if and only if the function $f$ is discrete (anti-)holomorphic. 
Choosing the weight function $W$ suitably for the quad-graphs with orthogonal diagonals, we have 
the topological quantum number
\begin{equation}
{\cal A}^{W}_{\diamondsuit}(f)=4 \pi N, \quad N \in \pi_2(S^2) \in \mathbb{Z}, 
\label{eq_DArea}
\end{equation}
through a kind of tiling of $S^2$ by the inverse of the stereographic projection. This ensures 
the topological stability of solutions to the Euler--Lagrange equation which is derived from 
the weighted discrete Dirichlet energy $E^{W}_{\diamondsuit}(f)$. 

The rest of this paper is organized as follows. 
In Section 2, we briefly summarize the discrete complex analysis by Mercat \cite{Mer01, Mer07}, Bobenko, 
and G{\"u}nther \cite{BobeGunt16}. In Section 3, a weighted discrete Dirichlet energy and area are defined, 
and an inequality relating the two is derived. This inequality is saturated if and only if the function $f$ is discrete (anti-)holomorphic. 
By defining the weight function $W$ adequately (see (\ref{eq_def_W(Q)}))
for the quad-graphs with orthogonal diagonals, 
we have a kind of tiling of 
the sphere by the inverse of the stereographic projection and the weighted discrete area admits 
a geometrical interpretation, i.e., a topological quantum number.
In Section 4, we derive the weighted discrete Euler--Lagrange equation and prove that the discrete (anti-)holomorphic function satisfies this equation. 
In Section 5, we adopt a rhombic lattice. 
We explicitly compute the topological number $N$ for the discrete power functions on the lattice. Moreover, we show that the weighted discrete Dirichlet energy, area, and Euler--Lagrange equation tend toward their continuous forms as the lattice spacings tend to zero. 
In Section 6, we present our final remarks.

\section{Discrete complex analysis of Mercat, Bobenko, and G{\"u}nther}

Here, we briefly present some definitions and formulas used by Mercat \cite{Mer01, Mer07} and Bobenko and G{\"u}nther \cite{BobeGunt16}. These definitions are employed in Section 3 to define the weighted discrete Dirichlet energy and area. Throughout this paper, we follow the notation of \cite{BobeGunt16}, unless otherwise stated.

Figure \ref{fig:Fv+FQ} shows a planar quad-graph $\Lambda$ (solid lines, black and white vertices) in which all edges are straight-line segments and all faces are convex or non-convex quadrilaterals, as well as the medial graph $X$ (thin lines and gray vertices) of $\Lambda$, with the vertices of $X$ defined as the midpoints of the edges of $\Lambda$. By connecting adjacent vertices of the face $F$ of the medial graph $X$, we obtain the parallelogram $F_{Q}$ (due to Varignon's theorem) in the quadrilateral $Q \in F(\Lambda)$ and the polygon $F_{v}$ corresponding to the vertex $v \in V(\Lambda)$. These two types of faces are respectively shown by hatching in Figures \ref{fig:Fv+FQ} (a) and (b).  
\begin{figure}[htbp]
\hspace{.5cm}
  \begin{minipage}[b]{0.48\linewidth}
    \centering
    \vspace{0ex}\hspace{0ex}\scalebox{0.9}{\input{Fv+FQ_gokusaisen_hatched_dense_1_tado.tex}}
    \subcaption{ }\label{fig:Fv+FQ_1}
  \end{minipage}
\hspace{-1cm}
  \begin{minipage}[b]{0.48\linewidth}
    \centering
    \vspace{0ex}\hspace{0ex}\scalebox{0.9}{\input{Fv+FQ_gokusaisen_hatched_dense_2_tado.tex}} 
    \subcaption{ }\label{fig:Fv+FQ_2}
  \end{minipage}
   \vspace*{-0ex}\hspace*{-3ex}\caption{Quad-graph $\Lambda$ (solid lines, black and white vertices) 
and its medial graph $X$ (thin lines and gray vertices). 
(a) Parallelogram $F_{Q}$ (blue) and (b) Polygon $F_{v}$ (red)}  
\label{fig:Fv+FQ}
\end{figure}
The dual graph of $\Lambda$ is denoted by $\diamondsuit:=\Lambda^{*}$. We have a bijection between the set $F(X)$ of faces of $X$ and the union $V(\diamondsuit) \cup V(\Lambda)$, where $V(\diamondsuit) \cong F(\Lambda)$ and $V(\Lambda)$ are the sets of vertices of $\diamondsuit$ and $\Lambda$, respectively. Graphs $\Gamma$ and $\Gamma^{*}$ denote the graphs on the black and white vertices, respectively. We assume that the planar quad-graph $\Lambda$ is embedded into the complex plane $\mathbb{C}$ and that the vertices and oriented edges are identified with their corresponding complex values. In this paper, we also assume that the graphs are connected.       

A function $f: V(\Lambda) \to \mathbb{C}$ is said to be discrete holomorphic at $Q$ if and only if it satisfies the discrete Cauchy--Riemann equation 
\begin{equation}
\frac{f(b_{+})-f(b_{-})}{b_{+}-b_{-}}=\frac{f(w_{+})-f(w_{-})}{w_{+}-w_{-}}, 
\label{eq_def_dholo}
\end{equation}
in which the four vertices $b_{-}$, $w_{-}$, $b_{+}$, and $w_{+}$ of $Q$ are assumed to be ordered counterclockwise, where $b_{\pm} \in V(\Gamma)$ and $w_{\pm} \in V(\Gamma^{*})$. In this paper, we also define discrete anti-holomorphicity, which is not described explicitly in \cite{BobeGunt16}. A function $f$ is said to be discrete anti-holomorphic at $Q$ if and only if it satisfies the discrete anti-Cauchy--Riemann equation 
\begin{equation}
\frac{f(b_{+})-f(b_{-})}{\overline{b_{+}-b_{-}}}=\frac{f(w_{+})-f(w_{-})}{\overline{w_{+}-w_{-}}}. 
\label{eq_def_daholo}
\end{equation}
A function $f: V(\Lambda) \to \mathbb{C}$ that is separately constant on $V(\Gamma)$ and $V(\Gamma^{*})$ is said to be biconstant. Such biconstants are discrete holomorphic and anti-holomorphic at any $Q$. By introducing the quantity \cite{Mer01, BobeGunt16}
\begin{equation}
\rho_{Q}:=-i\frac{w_{+}-w_{-}}{b_{+}-b_{-}}, 
\label{eq_def_rho}
\end{equation}
(which Mercat called the discrete conformal structure), the discrete Cauchy and anti-Cauchy--Riemann equations (\ref{eq_def_dholo}) and (\ref{eq_def_daholo}) are equivalent to
\begin{equation}
f(w_{+})-f(w_{-})=i\rho_{Q}\left(f(b_{+})-f(b_{-})\right)
\label{eq_dholo_rho}
\end{equation}
and
\begin{equation}
f(w_{+})-f(w_{-})=-i\bar{\rho}_{Q}\left(f(b_{+})-f(b_{-})\right), 
\label{eq_daholo_rho}
\end{equation}
respectively. If two diagonals of $Q$ are orthogonal to each other, 
then $\rho_{Q}$ is real and positive: $\rho_{Q}=|w_{+}-w_{-}|/|b_{+}-b_{-}| >0$, and equations (\ref{eq_dholo_rho}) and (\ref{eq_daholo_rho}) are written as follows:
\begin{equation}
f(w_{+})-f(w_{-})=\pm i\rho_{Q}\left(f(b_{+})-f(b_{-})\right). 
\label{eq_pm_rho_dholo_adholo}
\end{equation}
In general, the linear function is discrete holomorphic for any graph and the quadratic function is discrete holomorphic only for parallelogram graphs.

The discrete derivatives $\partial_{\Lambda}f$ and $\bar{\partial}_{\Lambda}f$ are defined as
\begin{eqnarray}
\partial_{\Lambda}f(Q):&=&
\lambda_{Q}\frac{f(b_{+})-f(b_{-})}{b_{+}-b_{-}}+
\bar{\lambda}_{Q}\frac{f(w_{+})-f(w_{-})}{w_{+}-w_{-}}, 
\label{eq_partial_f_Q} \\
\bar{\partial}_{\Lambda}f(Q):&=&
\bar{\lambda}_{Q}\frac{f(b_{+})-f(b_{-})}{\overline{b_{+}-b_{-}}}+
\lambda_{Q}\frac{f(w_{+})-f(w_{-})}{\overline{w_{+}-w_{-}}}, 
\label{eq_bar{partial}_f_Q}
\end{eqnarray}
where $2\lambda_{Q}:={\rm exp}\left(-i(\varphi_{Q}-\frac{\pi}{2})\right)/\sin(\varphi_{Q})$ and $\varphi_{Q}$ is the angle ($0<\varphi_{Q}<\pi$) under which the two diagonals of $Q$ intersect (i.e., $\varphi_{Q}:={\rm arg}(w_{+}-w_{-})-{\rm arg}(b_{+}-b_{-})$). If $\varphi_{Q}=\pi/2$, then $\lambda_{Q}=1/2$. This case is studied by Chelkak and Smirnov \cite{ChelkSmir11} for isoradial graphs.  
In a continuous limit in which the black diagonal $b_{+}-b_{-}$ is parallel to the real axis and $\varphi_{Q}=\pi/2$, the discrete derivatives (\ref{eq_partial_f_Q}) and (\ref{eq_bar{partial}_f_Q}) tend to $\frac{1}{2}(\partial_{x}-i\partial_{y})f=\partial_{z}f$ and $\frac{1}{2}(\partial_{x}+i\partial_{y})f=\partial_{\bar{z}}f$, respectively. As proved in \cite{BobeGunt16}, $f$ is discrete holomorphic at $Q$ if and only if $\bar{\partial}_{\Lambda}f(Q)=0$. We also have its anti-holomorphic version; $f$ is discrete anti-holomorphic at $Q$ if and only if $\partial_{\Lambda}f(Q)=0$. 

The discrete exterior derivative $df$ for the function $f: V(\Lambda) \to \mathbb{C}$ is defined as the discrete one-form on the oriented edges of $X$ given by
\begin{equation}
df:=\partial_{\Lambda}fdz+\bar{\partial}_{\Lambda}fd\bar{z}. 
\label{eq_df_BG}
\end{equation}
By integrating both sides of (\ref{eq_df_BG}) along the oriented edge $e$ of $X$ starting at the midpoint of the edge $vv'_{-}$ and ending at the midpoint of the edge $vv'_{+}$ of $\Lambda$, we have the discrete Stokes' theorem \cite{BobeGunt16}: 
\begin{equation}
\int_e df=\frac{f(v'_{+})-f(v'_{-})}{2}=\frac{f(v)+f(v'_{+})}{2}-\frac{f(v)+f(v'_{-})}{2}. 
\label{eq_Stokes_BG}
\end{equation}
If the function $f(z)$ is linear in $z$, then (\ref{eq_Stokes_BG}) gives $\int_e df=f(e_{+})-f(e_{-})$ with edge $e$ having terminal point $e_{+}:=(v+v'_{+})/2$ and initial point $e_{-}:=(v+v'_{-})/2$. By contrast, Mercat \cite{Mer01, Mer07} defined the discrete exterior derivative such that the discrete Stokes' theorem holds. 

The discrete Hodge star $\star$ is defined by 
\begin{eqnarray}
& & \star f:=-\frac{1}{2i}f\Omega_{\Lambda}, \quad \star h:=-\frac{1}{2i}h\Omega_{\diamondsuit}, 
\label{eq_*f_*h} \\ 
& & \star \omega=\star(pdz+qd\bar{z}):=-ipdz+iqd\bar{z}, 
\label{eq_*omega} \\
& & \star \Omega_{1}:=-2i\frac{\Omega_{1}}{\Omega_{\Lambda}}, 
\quad \star \Omega_{2}:=-2i\frac{\Omega_{2}}{\Omega_{\diamondsuit}}, \label{eq_*Omega1_*Omega2}
\end{eqnarray}
where $f$ and $h$ are functions of type $\Lambda$ and $\diamondsuit$, respectively, $\omega=pdz+qd\bar{z}$ is a discrete one-form of type $\diamondsuit$ defined on the oriented edges of $X$ with $p,\, q: V(\diamondsuit) \to \mathbb{C}$. Moreover, $\Omega_{1}$ and $\Omega_{2}$ are discrete two-forms of type $\Lambda$ and $\diamondsuit$ defined on $F_{Q}$ and $F_{v}$, respectively. The discrete two-forms $\Omega_{\Lambda}$ and $\Omega_{\diamondsuit}$ of type $\Lambda$ and $\diamondsuit$ are normalized as follows:
\begin{equation}
\iint_{F_v} \Omega_{\Lambda}=-4i\,{\rm area}(F_v), \quad 
\iint_{F_Q}\Omega_{\diamondsuit}=-4i\,{\rm area}(F_Q), 
\label{eq_int_Omega}
\end{equation}
respectively.

To express $\int_{e} \star \omega$ and $\int_{e^{*}} \star \omega$ in terms of $\int_{e} \omega$ and $\int_{e^{*}}\omega$, we consider the system of linear equations $\int_{e} \omega=pe+q\bar{e}$ and $\int_{e^{*}}\omega=pe^{*}+q\bar{e^{*}}$, where $e$ and $e^{*}$ are the oriented edges of $X$ parallel to the black and white diagonals of $Q$, respectively, such that ${\rm Im}(e^{*}/ e)>0$. 
The system of linear equations is nondegenerate; its determinant is given by $e\bar{e}^{*}-\bar{e}{e}^{*}=-i\,{\rm area}(Q) \ne 0$, and we have $p=\frac{\lambda_{Q}}{e}\int_{e} \omega+\frac{\bar{\lambda}_{Q}}{e^{*}}\int_{e^{*}}\omega$ and $q=\frac{\bar{\lambda}_{Q}}{\bar{e}}\int_{e} \omega+\frac{\lambda_{Q}}{\bar{e}^{*}}\int_{e^{*}}\omega$. 
Using (\ref{eq_*omega}), we have the following formulas: 
\begin{eqnarray}
\int_{e} \star \omega&=&\cot(\varphi_{Q})\int_{e} \omega-\frac{|e|}{|e^{*}|\sin(\varphi_{Q})} \int_{e^{*}} \omega, \label{eq_int_Hodge_omega_1} \\
\int_{e^{*}} \star \omega&=&\frac{|e^{*}|}{|e|\sin(\varphi_{Q})} \int_{e}\omega-\cot(\varphi_{Q})\int_{e^{*}} \omega. 
\label{eq_int_Hodge_omega_2} 
\end{eqnarray}
For example, from (\ref{eq_int_Hodge_omega_1}), (\ref{eq_int_Hodge_omega_2}), and (\ref{eq_Stokes_BG}), we obtain the following relations:
\begin{equation}
\int_{e} (df-i\star df)=2\bar{e}\bar{\partial}_{\Lambda}f(Q), \quad  
\int_{e} (df+i\star df)=2e\partial_{\Lambda}f(Q). 
\label{eq_int_df+-stardf=Q} 
\end{equation}
Because the edge $e$ in (\ref{eq_int_df+-stardf=Q}) is arbitrary, we see that the equations $df-i\star df=0$ and $df+i\star df=0$ are equivalent to $\bar{\partial}_{\Lambda}f(Q)=0$ and $\partial_{\Lambda}f(Q)=0$, respectively.

The products $f\omega$, $h\omega$, $f\Omega_{1}$, and $h\Omega_{2}$ are defined by
\begin{eqnarray}
\fl \hspace{.5cm} \int_e f\omega:&=&f(v)\int_e \omega, \quad\, \iint_{F_v} f\Omega_{1}:=f(v)\iint_{F_v}\Omega_{1}, \quad \iint_{F_Q} f\Omega_{1}:=0, 
\label{eq_products_f_omega_1} \\
\fl \hspace{.5cm} \int_e h\omega:&=&h(Q)\int_e \omega, \quad \iint_{F_v} h\Omega_{2}:=0, \hspace{2.4cm}
\iint_{F_Q} h\Omega_{2}:=h(Q)\iint_{F_Q}\Omega_{2}, 
\label{eq_products_h_omega_2}
\end{eqnarray}
where $e$ is the oriented edge of $X$, which is uniquely determined by the quadrilateral $Q$ and its vertex $v$ \cite{BobeGunt16}. To illustrate these formulas, integrating both sides of the equations of (\ref{eq_*f_*h}) (for example, over $F_v$ and $F_Q$, respectively), we have 
\begin{equation}
\iint_{F_v}\star f=2\,{\rm area}(F_v)f(v), \quad \iint_{F_Q}\star h=2\,{\rm area}(F_Q)h(Q). 
\label{eq_int_*f_*h}
\end{equation}
Similarly, from (\ref{eq_*Omega1_*Omega2}) with denominators being cleared, (\ref{eq_products_f_omega_1}), and (\ref{eq_products_h_omega_2}), we obtain 
\begin{equation}
(\star \Omega_{1})(v)=\frac{1}{2\,{\rm area}(F_v)}\iint_{F_v}\Omega_{1}, \quad 
(\star \Omega_{2})(Q)=\frac{1}{2\,{\rm area}(F_Q)}\iint_{F_Q}\Omega_{2}. 
\label{eq_int_*Omega}
\end{equation}

The discrete scalar products are defined as 
\begin{eqnarray}
\fl \qquad \langle f_{1}, f_{2} \rangle:=\iint_{F(X)} f_{1}\star\bar{f}_{2}=\sum_{v \in V(\Lambda)}
2\,{\rm area}(F_v)f_{1}(v)\bar{f}_{2}(v), \label{eq_<f1f2>} \\
\fl \qquad \langle h_{1}, h_{2} \rangle:=\iint_{F(X)} h_{1}\star\bar{h}_{2}=\sum_{Q \in V(\diamondsuit)}
2\,{\rm area}(F_Q)h_{1}(Q)\bar{h}_{2}(Q), \label{eq_<h1h2>} \\
\fl \qquad \langle \omega, \omega' \rangle:=\iint_{F(X)} \omega \wedge \star\bar{\omega}',  \label{eq_<omegaomega'>} \\ 
\fl \qquad \langle \Omega_{1}, \Omega'_{1} \rangle:=\iint_{F(X)} \Omega_{1} \star\bar{\Omega}'_{1}
=\sum_{v \in V(\Lambda)}\frac{1}{2\,{\rm area}(F_v)}\iint_{F_{v}}\Omega_{1}\cdot \iint_{F_{v}} \bar{\Omega}'_{1},  \label{eq_<Omega1Omega1'>} \\
\fl \qquad \langle \Omega_{2}, \Omega'_{2} \rangle:=\iint_{F(X)} \Omega_{2} \star\bar{\Omega}'_{2}
=\sum_{Q \in V(\diamondsuit)}\frac{1}{2\,{\rm area}(F_Q)}\iint_{F_Q}\Omega_{2}\cdot \iint_{F_{Q}} \bar{\Omega}'_{2}. 
\label{eq_<Omega2Omega2'>}
\end{eqnarray}
The middle parts of the equations (\ref{eq_<f1f2>}), (\ref{eq_<h1h2>}), (\ref{eq_<Omega1Omega1'>}), and (\ref{eq_<Omega2Omega2'>}) have been further expressed as sums over the vertex $v \in V(\Lambda)$ or the quadrilateral $Q \in V(\diamondsuit)$. The discrete scalar products defined above are Hermitian because they satisfy $\overline{\langle f_{1}, f_{2} \rangle} =\langle f_{2}, f_{1} \rangle$, etc., as immediately seen by their right-hand sides. The equations (\ref{eq_int_df+-stardf=Q}), (\ref{eq_int_*f_*h}), and (\ref{eq_int_*Omega}) and the right-hand sides of (\ref{eq_<f1f2>}), (\ref{eq_<h1h2>}), (\ref{eq_<Omega1Omega1'>}), and (\ref{eq_<Omega2Omega2'>}) are not stated explicitly in \cite{BobeGunt16}.  

The integral of the discrete wedge product $\omega \wedge \omega'$ is given by 
\begin{equation}
\iint_{F_Q}\omega \wedge \omega'=2\int_{e}\omega\int_{e^{*}}\omega'-2
\int_{e^{*}}\omega\int_{e}\omega'. 
\label{eq_omega_wedge_omega'}
\end{equation}

\section{The weighted discrete Dirichlet energy and the area on quad-graphs with orthogonal diagonals}

For any planar quad-graph $\Lambda$, we consider the inequality 
\begin{equation}
{\|}h(df \mp i \star df){\|}^2 \ge 0, 
\label{eq_norm^2}
\end{equation}
where $f$ and $h$ are functions of type $\Lambda$ and $\diamondsuit$, respectively, and the norm for the one-form $\omega$ of type $\diamondsuit$ is defined as ${\|} \omega {\|}^2:= \langle \omega, \omega \rangle$. 
From (\ref{eq_norm^2}), we immediately obtain a discrete version of the inequality (\ref{eq_E_ineq}) 
\begin{equation}
E^{W}_{\diamondsuit}(f)
= {\|}h(df \mp i \star df){\|}^2
\pm {\cal A}^{W}_{\diamondsuit}(f) \ge | {\cal A}^{W}_{\diamondsuit}(f)|, 
\label{eq_Dineq}
\end{equation}
where the weighted discrete Dirichlet energy $E^{W}_{\diamondsuit}(f)$ and weighted discrete area ${\cal A}^{W}_{\diamondsuit}(f)$ of $f$ on $\diamondsuit$ are defined as
\begin{eqnarray}
\fl \hspace{0.8cm} E^{W}_{\diamondsuit}(f)
:&=&2\langle hdf, hdf \rangle =2\langle Wdf, df \rangle \ge 0, 
\label{eq_def_WDEnergy} \\
\fl \hspace{0.8cm} {\cal A}^{W}_{\diamondsuit}(f)
:&=&-2i\langle hdf, h\star df \rangle =-2i\langle Wdf, \star \, df \rangle.
\label{eq_def_WDArea}
\end{eqnarray}
Here we call $W\!:=|h|^{2}$ the weight or weight function. 
We have $\langle hdf, hdf \rangle =\langle h\star\hspace{.4mm} df, h\star\hspace{.4mm} df \rangle$, and moreover $\langle hdf, h\star df \rangle$ is pure imaginary, because we have $\star^{2}=-{\rm Id}$ on discrete one-forms of type $\diamondsuit$ defined on oriented edges of $X$. The inequality (\ref{eq_Dineq}) is saturated if and only if the function $f$ is discrete (anti-)holomorphic: $df \mp i\star df=0$, provided that $h$ is nowhere vanishing. 

From this point onward, we restrict ourselves to quad-graphs with orthogonal diagonals such as the rhombic lattice in Sections 5 and the isoradial graphs in \cite{ChelkSmir11}. The discrete conformal structure $\rho_{Q}$ is real and positive for these graphs. 
Using the formulas presented in the preceding section, more explicit forms of (\ref{eq_norm^2}), (\ref{eq_def_WDEnergy}), and (\ref{eq_def_WDArea}) are given as
\begin{eqnarray}
\fl {\|}h(df \mp i \star df){\|}^2
&=&\!\!\sum_{Q \in V(\diamondsuit)}W
\biggl|\sqrt{\rho^{ }_{Q}}(f(b_{+})-f(b_{-})) \pm i\sqrt{\rho^{-1}_{Q}}(f(w_{+})-f(w_{-}))\biggr|^2 
\ge 0, \hspace{.7cm} 
\label{eq_norm^2_comp} \\ 
\fl \hspace{0.8cm} E^{W}_{\diamondsuit}(f)
&=&\!\!\sum_{Q \in V(\diamondsuit)}W\Bigl(\rho_{Q}\bigl|f(b_{+})-f(b_{-})\bigr|^2+
\rho^{-1}_{Q}\bigl|f(w_{+})-f(w_{-})\bigr|^2\Bigr) \ge 0, 
\label{eq_WDEnergy_comp} \\
\fl \hspace{0.8cm} {\cal A}^{W}_{\diamondsuit}(f)
&=&2\,{\rm Im}\left(\sum_{Q \in V(\diamondsuit)}W(f(w_{+})-f(w_{-}))\overline{(f(b_{+})-f(b_{-}))}\right), 
\label{eq_WDArea_comp}
\end{eqnarray}
respectively. The weighted discrete Dirichlet energy (\ref{eq_WDEnergy_comp}) is positive semi-definite and depends upon $\rho_{Q}$, whereas the weighted discrete area (\ref{eq_WDArea_comp}) does not have these properties. 

If $W=1$, then the weighted discrete Dirichlet energy (\ref{eq_def_WDEnergy}) is twice the discrete Dirichlet energy $E_{\diamondsuit}(f)=\langle df, df \rangle$ of \cite{BobeGunt16}. On the other hand, we impose the condition 
\begin{equation}
W \to (1+|f(z)|^2)^{-2} \quad {\rm as} \quad Q \to {\rm a\,\, single\,\, point}\,\, z.  
\label{eq_W(Q)_limt_z}
\end{equation}
This condition indeed holds for the rhombic lattice in Section 5, see (\ref{eq_WQ1}).  

Now, we rewrite the weighted discrete area (\ref{eq_WDArea_comp}) as 
\begin{equation}
\fl \hspace{1cm} {\cal A}^{W}_{\diamondsuit}(f) =4\sum_{Q \in V(\diamondsuit)}W{\rm area}([f(Q)]), 
\label{eq_4SumWarea} 
\end{equation}
where 
\begin{equation}
\fl \hspace{0.8cm} {\rm area}([f(Q)]):=
\frac{1}{2}\,{\rm Im}\left((f(w_{+})-f(w_{-}))\overline{(f(b_{+})-f(b_{-}))}\right) 
\label{eq_signed_area} 
\end{equation}
is a {\it signed} area of the quadrilateral consisting of the four ordered vertices $f(b_{-})$, $f(w_{-})$, $f(b_{+})$, and $f(w_{+})$ with edges given by straight-line segments. Hereafter in this paper, this quadrilateral is denoted by the symbol $[f(Q)]$ (Figure \ref{fig:[f(Q)]}). 
\begin{figure}[htbp]
  \begin{center}
   \vspace{1ex}\scalebox{1.0}{{\unitlength=0.66667cm%
\begin{picture}%
(5.80,5.90)(-0.30,-1.30)%
\special{pn 8}%
\special{pa 0 0}\special{pa 1391 210}\special{pa 1312 -1050}\special{pa 394 -919}%
\special{pa 0 0}%
\special{fp}%
\special{pa 21 0}\special{pa 21 -3}\special{pa 20 -5}\special{pa 19 -8}\special{pa 18 -10}%
\special{pa 17 -12}\special{pa 15 -14}\special{pa 13 -16}\special{pa 11 -18}\special{pa 9 -19}%
\special{pa 6 -20}\special{pa 3 -21}\special{pa 1 -21}\special{pa -2 -21}\special{pa -4 -21}%
\special{pa -7 -20}\special{pa -10 -19}\special{pa -12 -17}\special{pa -14 -16}\special{pa -16 -14}%
\special{pa -18 -11}\special{pa -19 -9}\special{pa -20 -7}\special{pa -21 -4}\special{pa -21 -1}%
\special{pa -21 1}\special{pa -21 4}\special{pa -20 7}\special{pa -19 9}\special{pa -18 12}%
\special{pa -16 14}\special{pa -14 16}\special{pa -12 17}\special{pa -10 19}\special{pa -7 20}%
\special{pa -5 20}\special{pa -2 21}\special{pa 1 21}\special{pa 3 21}\special{pa 6 20}%
\special{pa 8 19}\special{pa 11 18}\special{pa 13 16}\special{pa 15 15}\special{pa 17 13}%
\special{pa 18 10}\special{pa 19 8}\special{pa 20 5}\special{pa 21 3}\special{pa 21 0}%
\special{fp}%
\special{pa 1412 210}\special{pa 1412 207}\special{pa 1411 205}\special{pa 1411 202}%
\special{pa 1409 200}\special{pa 1408 197}\special{pa 1406 195}\special{pa 1404 194}%
\special{pa 1402 192}\special{pa 1400 191}\special{pa 1397 190}\special{pa 1394 189}%
\special{pa 1392 189}\special{pa 1389 189}\special{pa 1387 189}\special{pa 1384 190}%
\special{pa 1381 191}\special{pa 1379 193}\special{pa 1377 194}\special{pa 1375 196}%
\special{pa 1373 199}\special{pa 1372 201}\special{pa 1371 203}\special{pa 1370 206}%
\special{pa 1370 209}\special{pa 1370 211}\special{pa 1370 214}\special{pa 1371 217}%
\special{pa 1372 219}\special{pa 1374 222}\special{pa 1375 224}\special{pa 1377 226}%
\special{pa 1379 227}\special{pa 1381 229}\special{pa 1384 230}\special{pa 1386 230}%
\special{pa 1389 231}\special{pa 1392 231}\special{pa 1394 231}\special{pa 1397 230}%
\special{pa 1399 229}\special{pa 1402 228}\special{pa 1404 226}\special{pa 1406 225}%
\special{pa 1408 223}\special{pa 1409 220}\special{pa 1410 218}\special{pa 1411 215}%
\special{pa 1412 213}\special{pa 1412 210}%
\special{fp}%
\special{pa 1333 -1050}\special{pa 1333 -1052}\special{pa 1333 -1055}\special{pa 1332 -1058}%
\special{pa 1331 -1060}\special{pa 1329 -1062}\special{pa 1328 -1064}\special{pa 1325 -1066}%
\special{pa 1323 -1068}\special{pa 1321 -1069}\special{pa 1318 -1070}\special{pa 1316 -1071}%
\special{pa 1313 -1071}\special{pa 1310 -1071}\special{pa 1308 -1070}\special{pa 1305 -1070}%
\special{pa 1303 -1068}\special{pa 1300 -1067}\special{pa 1298 -1065}\special{pa 1296 -1064}%
\special{pa 1295 -1061}\special{pa 1293 -1059}\special{pa 1292 -1056}\special{pa 1292 -1054}%
\special{pa 1291 -1051}\special{pa 1291 -1049}\special{pa 1292 -1046}\special{pa 1292 -1043}%
\special{pa 1293 -1041}\special{pa 1295 -1038}\special{pa 1296 -1036}\special{pa 1298 -1034}%
\special{pa 1300 -1033}\special{pa 1303 -1031}\special{pa 1305 -1030}\special{pa 1308 -1029}%
\special{pa 1310 -1029}\special{pa 1313 -1029}\special{pa 1315 -1029}\special{pa 1318 -1030}%
\special{pa 1321 -1031}\special{pa 1323 -1032}\special{pa 1325 -1033}\special{pa 1327 -1035}%
\special{pa 1329 -1037}\special{pa 1331 -1040}\special{pa 1332 -1042}\special{pa 1333 -1044}%
\special{pa 1333 -1047}\special{pa 1333 -1050}%
\special{fp}%
\special{pa 415 -919}\special{pa 415 -921}\special{pa 414 -924}\special{pa 413 -926}%
\special{pa 412 -929}\special{pa 411 -931}\special{pa 409 -933}\special{pa 407 -935}%
\special{pa 405 -937}\special{pa 402 -938}\special{pa 400 -939}\special{pa 397 -939}%
\special{pa 395 -940}\special{pa 392 -940}\special{pa 389 -939}\special{pa 387 -938}%
\special{pa 384 -937}\special{pa 382 -936}\special{pa 380 -934}\special{pa 378 -932}%
\special{pa 376 -930}\special{pa 375 -928}\special{pa 374 -925}\special{pa 373 -923}%
\special{pa 373 -920}\special{pa 373 -917}\special{pa 373 -915}\special{pa 374 -912}%
\special{pa 375 -909}\special{pa 376 -907}\special{pa 378 -905}\special{pa 380 -903}%
\special{pa 382 -901}\special{pa 384 -900}\special{pa 386 -899}\special{pa 389 -898}%
\special{pa 392 -898}\special{pa 394 -898}\special{pa 397 -898}\special{pa 400 -898}%
\special{pa 402 -899}\special{pa 405 -901}\special{pa 407 -902}\special{pa 409 -904}%
\special{pa 411 -906}\special{pa 412 -908}\special{pa 413 -911}\special{pa 414 -913}%
\special{pa 414 -916}\special{pa 415 -919}%
\special{fp}%
\settowidth{\Width}{$f(b_-)$}\setlength{\Width}{-1.\Width}%
\settoheight{\Height}{$f(b_-)$}\settodepth{\Depth}{$f(b_-)$}\setlength{\Height}{-\Height}%
\put(-0.0750,-0.0750){\hspace*{\Width}\raisebox{\Height}{$f(b_-)$}}%
\settowidth{\Width}{$f(w_-)$}\setlength{\Width}{0.\Width}%
\settoheight{\Height}{$f(w_-)$}\settodepth{\Depth}{$f(w_-)$}\setlength{\Height}{-\Height}%
\put(5.3750,-0.8750){\hspace*{\Width}\raisebox{\Height}{$f(w_-)$}}%
\settowidth{\Width}{$f(b_+)$}\setlength{\Width}{0.\Width}%
\settoheight{\Height}{$f(b_+)$}\settodepth{\Depth}{$f(b_+)$}\setlength{\Height}{\Depth}%
\put(5.0750,4.0750){\hspace*{\Width}\raisebox{\Height}{$f(b_+)$}}%
\settowidth{\Width}{$f(w_+)$}\setlength{\Width}{-1.\Width}%
\settoheight{\Height}{$f(w_+)$}\settodepth{\Depth}{$f(w_+)$}\setlength{\Height}{\Depth}%
\put(1.4250,3.5750){\hspace*{\Width}\raisebox{\Height}{$f(w_+)$}}%
\settowidth{\Width}{$[f(Q)]$}\setlength{\Width}{-0.5\Width}%
\settoheight{\Height}{$[f(Q)]$}\settodepth{\Depth}{$[f(Q)]$}\setlength{\Height}{-0.5\Height}\setlength{\Depth}{0.5\Depth}\addtolength{\Height}{\Depth}%
\put(2.9500,1.6750){\hspace*{\Width}\raisebox{\Height}{$[f(Q)]$}}%
\special{pa 21 0}\special{pa 20 -5}\special{pa 18 -10}\special{pa 17 -12}\special{pa 13 -16}%
\special{pa 11 -18}\special{pa 6 -20}\special{pa 1 -21}\special{pa -2 -21}\special{pa -7 -20}%
\special{pa -10 -19}\special{pa -14 -16}\special{pa -18 -11}\special{pa -19 -9}\special{pa -21 -4}%
\special{pa -21 1}\special{pa -21 4}\special{pa -19 9}\special{pa -18 12}\special{pa -14 16}%
\special{pa -10 19}\special{pa -7 20}\special{pa -2 21}\special{pa 3 21}\special{pa 6 20}%
\special{pa 11 18}\special{pa 13 16}\special{pa 17 13}\special{pa 19 8}\special{pa 20 5}%
\special{pa 21 0}\special{sh 1}\special{ip}%
\special{pa 1412 210}\special{pa 1411 205}\special{pa 1409 200}\special{pa 1408 197}%
\special{pa 1404 194}\special{pa 1402 192}\special{pa 1397 190}\special{pa 1392 189}%
\special{pa 1389 189}\special{pa 1384 190}\special{pa 1381 191}\special{pa 1377 194}%
\special{pa 1373 199}\special{pa 1372 201}\special{pa 1370 206}\special{pa 1370 211}%
\special{pa 1370 214}\special{pa 1372 219}\special{pa 1374 222}\special{pa 1377 226}%
\special{pa 1381 229}\special{pa 1384 230}\special{pa 1389 231}\special{pa 1394 231}%
\special{pa 1397 230}\special{pa 1402 228}\special{pa 1404 226}\special{pa 1408 223}%
\special{pa 1410 218}\special{pa 1411 215}\special{pa 1412 210}\special{sh 0}\special{ip}%
\special{pa 1333 -1050}\special{pa 1333 -1055}\special{pa 1331 -1060}\special{pa 1329 -1062}%
\special{pa 1325 -1066}\special{pa 1323 -1068}\special{pa 1318 -1070}\special{pa 1313 -1071}%
\special{pa 1310 -1071}\special{pa 1305 -1070}\special{pa 1303 -1068}\special{pa 1298 -1065}%
\special{pa 1295 -1061}\special{pa 1293 -1059}\special{pa 1292 -1054}\special{pa 1291 -1049}%
\special{pa 1292 -1046}\special{pa 1293 -1041}\special{pa 1295 -1038}\special{pa 1298 -1034}%
\special{pa 1303 -1031}\special{pa 1305 -1030}\special{pa 1310 -1029}\special{pa 1315 -1029}%
\special{pa 1318 -1030}\special{pa 1323 -1032}\special{pa 1325 -1033}\special{pa 1329 -1037}%
\special{pa 1332 -1042}\special{pa 1333 -1044}\special{pa 1333 -1050}\special{sh 1}\special{ip}%
\special{pa 415 -919}\special{pa 414 -924}\special{pa 412 -929}\special{pa 411 -931}%
\special{pa 407 -935}\special{pa 405 -937}\special{pa 400 -939}\special{pa 395 -940}%
\special{pa 392 -940}\special{pa 387 -938}\special{pa 384 -937}\special{pa 380 -934}%
\special{pa 376 -930}\special{pa 375 -928}\special{pa 373 -923}\special{pa 373 -917}%
\special{pa 373 -915}\special{pa 375 -909}\special{pa 376 -907}\special{pa 380 -903}%
\special{pa 384 -900}\special{pa 386 -899}\special{pa 392 -898}\special{pa 397 -898}%
\special{pa 400 -898}\special{pa 405 -901}\special{pa 407 -902}\special{pa 411 -906}%
\special{pa 413 -911}\special{pa 414 -913}\special{pa 415 -919}\special{sh 0}\special{ip}%
\end{picture}}%
} \qquad
   \vspace*{-0ex}\hspace*{-3ex}\caption{A quadrilateral $[f(Q)]$ and its vertices} 
\label{fig:[f(Q)]}
  \end{center} 
\end{figure}
There exist several types of quadrilaterals $[f(Q)]$, namely, simple (i.e., their edges do not cross each other) convex or non-convex quadrilaterals and non-simple ones. The latter may include some degenerate cases such as bowtie-shaped quadrilaterals or even a single point, depending on the function $f$.  
If the function $f$ is the identity map, then (\ref{eq_signed_area}) reduces to the formula for the area of the quadrilateral $Q$ itself
\begin{equation}
{\rm area}(Q)=\frac{1}{2}\,{\rm Im}\left((w_{+}-w_{-})\overline{(b_{+}-b_{-})}\right)>0.  
\label{eq_area_Q} 
\end{equation}
This formula is valid for the quadrilateral $Q$ whose diagonals are not necessarily orthogonal. 
Note that (\ref{eq_area_Q}) is always positive because the four vertices $b_{-}$, $w_{-}$, $b_{+}$, and $w_{+}$ of the quadrilateral $Q$ are assumed to be ordered counterclockwise in Section 2. 

In order to define a topological quantum number, it is appropriate to define the weight function $W$ as 
\begin{equation}
W:=\frac{1}{4}\frac{{\rm area}({\rm St}^{-1}([f(Q)]))}{{\rm area}([f(Q)])}, \quad 
{\rm area}([f(Q)]) \ne 0
\label{eq_def_W(Q)}
\end{equation}
in accordance with the condition (\ref{eq_W(Q)_limt_z}). The value of $W$ at ${\rm area}([f(Q)]) = 0$ is arbitrary because it does not contribute to the weighted discrete area (\ref{eq_4SumWarea}).  
Here, ${\rm St^{-1}}$ is the inverse of the stereographic projection ${\rm St}: S^2 \rightarrow \mathbb{R}^2$. The inverse ${\rm St^{-1}}$ maps a region on $\mathbb{R}^2$ to the corresponding spherical one on $S^2$. 
Hereafter the weight function is written as $W([f(Q)])$.
From (\ref{eq_4SumWarea}) and (\ref{eq_def_W(Q)}), we obtain the formula  
\begin{equation}
{\cal A}^{W}_{\diamondsuit}(f) =\hspace{-1cm} \sum_{\hspace{1cm} Q \in V(\diamondsuit), \,\,\,{\rm area}([f(Q)]) \ne 0}\hspace{-1cm} {\rm area}({\rm St}^{-1}([f(Q)])), 
\label{eq_Sum_area} 
\end{equation}
which defines a kind of tiling of the unit sphere $S^2$ by the inverse stereographic projection. 

If the function $f$ is discrete (anti-)holomorphic at $Q$, 
whose diagonals may or may not be orthogonal, 
then we obtain 
\begin{equation}
{\rm area}([f(Q)]) \, \cases
{
\ge 0 & if the $f$ is discrete holomorphic, \\ 
\le 0 & if the $f$ is discrete anti-holomorphic, 
} 
\label{eq_def_area>=<0}
\end{equation}
from the discrete Cauchy--Riemann equation (\ref{eq_def_dholo}), its anti-version (\ref{eq_def_daholo}), and the inequality ${\rm Re}(\rho_{Q})>0$. 
The zero-area case of (\ref{eq_def_area>=<0}) gives the following equivalences
\begin{equation}
{\rm area}([f(Q)])=0 \,\,\, \Longleftrightarrow \,\,\, f(b_{+})=f(b_{-}) \,\,\, \Longleftrightarrow \,\,\, f(w_{+})=f(w_{-}).
\label{eq_def_area=0}
\end{equation}
The biconstant satisfies these relations. In this case, the quadrilateral $[f(Q)]$ becomes a line segment connecting the two points $f(b_{+})
$ and $f(w_{+})$ when $f(b_{+}) \ne f(w_{+})$, or a single point $f(b_{+})$ when $f(b_{+})=f(w_{+})$. 
Since these degenerate quadrilaterals yield vanishing areas on the sphere $S^2$, 
the formula (\ref{eq_Sum_area}) can be written in a simpler form 
\begin{equation}
{\cal A}^{W}_{\diamondsuit}(f) =\sum_{Q \in V(\diamondsuit)}{\rm area}({\rm St}^{-1}([f(Q)])).
\label{eq_Sum_area_simpler} 
\end{equation}
Moreover, we comment on 
the relative angle $\varphi_{[f(Q)]}:={\rm arg}\{(f(w_{+})-f(w_{-}))/(f(b_{+})-f(b_{-}))\} \in (-\pi, \pi]$ between the two non-zero complex numbers $f(w_{+})-f(w_{-})$ and $f(b_{+})-f(b_{-})$. 
The discrete (anti-)Cauchy--Riemann equation 
(\ref{eq_def_dholo}) ((\ref{eq_def_daholo})) gives 
\begin{equation}
\varphi_{[f(Q)]} = \cases
{
\varphi_{Q} & if the $f$ is discrete holomorphic, \\  
-\varphi_{Q} & if the $f$ is discrete anti-holomorphic. 
} 
\label{eq_area_case}
\end{equation}
This shows that the relative angle 
is preserved up to orientation by the discrete (anti-)holomorphic function $f$ (See, p.64 of  \cite{BobeGunt16}). 

In Section 5, we will compute the weighted discrete area through the formula (\ref{eq_Sum_area_simpler}) for the discrete 
holomorphic power functions $z^{(N)}$ defined on a rhombic lattice 
and show the relation 
\begin{equation}
{\cal A}^{W}_{\diamondsuit}(f)=4 \pi N, \quad N \in \mathbb{Z}_{\ge 0}. 
\label{eq_4piN}
\end{equation}
The integer $N \in \pi_2(S^2)$, a topological quantum number, classifies the map from $\mathbb{R}^2$ or the Riemann sphere $\mathbb{C}\cup\{\infty\} \cong S^2$ to $S^2$. 
This ensures the topological stability of the solutions described by the 
functions $z^{(N)}$, and we have 
the quantized energy $E^{W}_{\diamondsuit}(f)={\cal A}^{W}_{\diamondsuit}(f)=4 \pi N$.

If the weighted discrete area (\ref{eq_Sum_area_simpler}) is a topological quantum number, we expect that the discrete (anti-)holomorphic function $f$ satisfies the Euler--Lagrange equation derived from the weighted discrete Dirichlet energy $E^{W}_{\diamondsuit}(f)$. More precisely, at the saturation of the inequality (\ref{eq_Dineq}), we have the equality $E^{W}_{\diamondsuit}(f)=\pm {\cal A}^{W}_{\diamondsuit}(f)$ with a discrete (anti-)holomorphic function $f$. The first variation of both sides of this equality yields $\delta E^{W}_{\diamondsuit}(f)=\pm \delta{\cal A}^{W}_{\diamondsuit}(f)$. However, it will be zero because ${\cal A}^{W}_{\diamondsuit}(f)$ is a topological quantum number. Thus, the stationary condition $\delta E^{W}_{\diamondsuit}(f)=0$ (or the Euler--Lagrange equation) is satisfied by the discrete (anti-)holomorphic function $f$. 

In the next section, we derive the Euler--Lagrange equation from the 
weighted discrete Dirichlet 
energy $E^{W}_{\diamondsuit}(f)$ defned on the quad-graphs with orthogonal diagonals and 
show that the discrete (anti-)holomorphic function $f$ satisfies this equation. 

\section{The weighted discrete Euler--Lagrange equation for quad-graphs with orthogonal diagonals}

To apply the standard variational calculus used in \cite{BobeGunt16} to the weighted discrete Dirichlet energy (\ref{eq_def_WDEnergy}), we first consider the following two partial derivatives:
\begin{equation}
\frac{\partial E^{W}_{\diamondsuit}}{\partial f_{1}(v_{0})}(f, \bar{f})
:=\frac{d}{dt}E^{W}_{\diamondsuit}(f_{1}+t\phi, f_{2}){\biggr|}_{t=0} 
\label{eq_partial1_E}  
\end{equation}
and
\begin{equation}
\frac{\partial E^{W}_{\diamondsuit}}{\partial f_{2}(v_{0})}(f, \bar{f})
:=\frac{d}{dt}E^{W}_{\diamondsuit}(f_{1}, f_{2}+t\phi){\biggr|}_{t=0}. 
\label{eq_partial2_E}
\end{equation}
Here, $f_{1}$ and $f_{2}$ are the real and imaginary parts of the function $f$, respectively, 
$v_{0}$ is a vertex of $V(\Lambda)$ (for finite $\Lambda$, $v_{0}$ is an interior vertex of  
$V(\Lambda)$), 
and $\phi(v):=\delta_{vv_{0}}$ is the Kronecker delta on $V(\Lambda)$. We often use notation like $E^{W}_{\diamondsuit}(f, \bar{f})=E^{W}_{\diamondsuit}(f_{1}, f_{2})$ when no confusion can arise. This notation is already used in (\ref{eq_partial1_E}) and (\ref{eq_partial2_E}).

We decompose the area of $Q \in F(\Lambda)$ 
into the sum of the ones of four triangles along the boundary of the $Q$: 
\begin{equation}
\hspace{.2cm} {\rm area}(Q) =\sum_{e \in E(\partial Q)}\mathrm{area}(\triangle Pe_{-}e_{+}). 
\label{eq_cycl_sum_Q}
\end{equation}
All areas here are signed ones, and the decomposition (\ref{eq_cycl_sum_Q}) is independent of the choice of $P$. Hereafter, we place the point $P$ at the coordinate origin, $O$. Moreover, we decompose the spherical area corresponding to the quadrilateral $[f(Q)]$ as follows:
\begin{equation}
{\rm area}({\rm St}^{-1}([f(Q)]))=\sum_{e \in E(\partial Q)}\mathrm{area}({\rm St}^{-1}(\triangle Of(e_{-})f(e_{+}))). \label{eq_cycl_sum_St_inv} 
\end{equation}
In particular, the spherical area corresponding to a triangle $\triangle Oz_{1}z_{2}$ (where $z_{1}$ and $z_{2}$ are two points on $\mathbb{C}$) is given by
\begin{equation}
\fl \hspace{0cm}{\rm area}({\rm St}^{-1}(\triangle{Oz_{1}z_{2}}))
=\frac{2{\rm Im}(z_{2}\bar{z_{1}})}
{\sqrt{|z_{2}-z_{1}|^{2}+[{\rm Im}(z_{2}\bar{z_{1}})]^{2}}} 
{\rm tan}^{-1}
\frac{\sqrt{|z_{2}-z_{1}|^{2}+[{\rm Im}(z_{2}\bar{z_{1}})]^{2}}} 
{1+{\rm Re}(z_{2}\bar{z_{1}})}. 
\label{eq_Oz1z2_Sec4}
\end{equation}
A derivation of (\ref{eq_Oz1z2_Sec4}) is given in Appendix A, where another derivation using the Gauss--Bonnet formula is also presented. 
\begin{figure}[htbp]
  \begin{center}
   \vspace{1ex}\scalebox{1.0}{\input{Fig3_tado.tex}} \qquad
   \vspace*{-0ex}\hspace*{-3ex}\caption{Quadrilateral $Q_{s}$ incident to the vertex $v_{0}$ and its vertices (in the case of black $v_{0}$)} 
\label{fig:Q_{s}_and_its_vertices}
  \end{center} 
\end{figure}
Let $Q_{s}$ be a quadrilateral incident to the vertex $v_{0}$, i.e., $Q_{s} \sim v_{0}$ (see Figure \ref{fig:Q_{s}_and_its_vertices}). 
From (\ref{eq_cycl_sum_St_inv}) and (\ref{eq_Oz1z2_Sec4}), we have  
\begin{equation}
\frac{d}{dt}{\rm area}({\rm St}^{-1}([f(Q_{s})])(f_{1}+t\phi, f_{2}){\biggr|}_{t=0}
=-(J_{s}-J_{s-1}), 
\label{eq_J-J} 
\end{equation}
where $J_{s}$ is defined by
\begin{equation}
J_{s}:=\frac{\partial F_{s}}{\partial f_{1}(v_{0})} \quad {\rm with} \quad F_{s}:=
{\rm area}({\rm St}^{-1}(\triangle{Of(v_{0})f(v'_{s})})). 
\label{eq_def_Js} 
\end{equation}
Notably, only the two terms of (\ref{eq_cycl_sum_St_inv}) that include the value $f(v_{0})$ survive in (\ref{eq_J-J}). Moreover, the relative minus sign in front of $J_{s-1}$ to $J_{s}$ in (\ref{eq_J-J}) stems from the reverse ordering of the vertices between the two relevant triangles, namely, 
\begin{equation}
\fl \hspace{1.5cm}{\rm area}({\rm St}^{-1}(\triangle{Of(v'_{s-1})f(v_{0})}))
=-{\rm area}({\rm St}^{-1}(\triangle{Of(v_{0})f(v'_{s-1})}))=-F_{s-1}. 
\label{eq_origin_minus_F} 
\end{equation}

Applying (\ref{eq_signed_area}) to $Q_{s}$, we have 
\begin{equation}
\frac{d}{dt}{\rm area}([f(Q_{s})])(f_{1}+t\phi, f_{2}){\biggr|}_{t=0}=-\frac{1}{2}\,{\rm Im}(f(v'_{s})-f(v'_{s-1})).  
\label{eq_ddt_area_f(Q)} 
\end{equation}
From (\ref{eq_def_W(Q)}), (\ref{eq_J-J}), and (\ref{eq_ddt_area_f(Q)}), we obtain
\begin{equation}
\fl \hspace{0cm} \frac{d}{dt}W([f(Q_{s})])(f_{1}+t\phi, f_{2}){\biggr|}_{t=0}
\!\!\!\!\!=-\frac{J_{s}-J_{s-1}}{4\,{\rm area}([f(Q_{s})])}
+\frac{W([f(Q_{s})])}{2\,{\rm area}([f(Q_{s})])}{\rm Im}(f(v'_{s})-f(v'_{s-1})).  
\label{eq_ddt_W(Q)} 
\end{equation}
This relation (\ref{eq_ddt_W(Q)}) and 
\begin{equation}
\frac{d}{dt}\bigl|f(v_{s})-f(v_{0})\bigr|^2(f_{1}+t\phi, f_{2}){\biggr|}_{t=0}=-2\,{\rm Re}(f(v_{s})-f(v_{0})) 
\label{eq_ddt_|f|_2} 
\end{equation}
yield
\begin{equation}
\fl \hspace{.5cm} \frac{\partial E^{W}_{\diamondsuit}}{\partial f_{1}(v_{0})}(f, \bar{f})
=-\sum_{Q_{s} \sim v_{0}}\frac{e_{\diamondsuit}([f(Q_{s})])}{4\,{\rm area}([f(Q_{s})])}(J_{s}-J_{s-1})
+\sum_{Q_{s} \sim v_{0}}\frac{W([f(Q_{s})])}{2\,{\rm area}([f(Q_{s})])}K_{s}. 
\label{eq_partial1_JK_s} 
\end{equation}
Here, $e_{\diamondsuit}([f(Q_{s})])$ and $K_{s}$ are defined by
\begin{equation}
e_{\diamondsuit}([f(Q_{s})]):=\rho_{s}\bigl|f(v_{s})-f(v_{0})\bigr|^2+\rho^{-1}_{s}\bigl|f(v'_{s})-f(v'_{s-1})\bigr|^2,   
\label{eq_e_diamond} 
\end{equation}
\begin{equation}
K_{s}:=8\,{\rm area}(Q_{s}){\rm Im}\left[ \overline{(f(v'_{s})-f(v'_{s-1}))}
\partial_{\Lambda}f(Q_{s})\bar{\partial}_{\Lambda}f(Q_{s}) \right], 
\label{eq_def_Ks} 
\end{equation}
and $\rho_{s}$ denotes $\rho_{Q_{s}}$ if $v$ is black and $\rho^{-1}_{Q_{s}}$ if $v$ is white 
\cite{BobeGunt16}. 
In deriving (\ref{eq_partial1_JK_s}), we made use of the identity
\begin{equation}
\fl \hspace{1cm} \left(f(v'_{s})-f(v'_{s-1})\right)^{2}+\rho^{2}_{s}\left(f(v_{s})-f(v_{0})\right)^{2}
=4|v'_{s}-v'_{s-1}|^{2}\,\partial_{\Lambda}f(Q_{s})\bar{\partial}_{\Lambda}f(Q_{s}), 
\label{eq_dholo_identity} 
\end{equation}
which is a direct result of (\ref{eq_partial_f_Q}) and (\ref{eq_bar{partial}_f_Q}) when $\varphi_{Q_{s}}=\pi/2$. In a similar way, an expression for another partial derivative (\ref{eq_partial2_E}) is obtained. Finally, the following weighted discrete Euler--Lagrange equation is obtained as
\begin{eqnarray}
\fl \hspace{0cm}[{\rm EL}]^{\rm disc.}(v):&=&-\frac{\partial E^{W}_{\diamondsuit}}{\partial \bar{f}(v)}(f, \bar{f}) 
\nonumber \\
\fl \hspace{0cm} &=&
\sum_{Q_{s} \sim v}\frac{e_{\diamondsuit}([f(Q_{s})])}{\,4{\rm area}([f(Q_{s})])} 
\left(\frac{\partial F_{s}\,\,\,\,\,}{\partial \bar{f}(v)}-\frac{\partial F_{s-1}}{\partial \bar{f}(v)}\right) 
\nonumber \\
\fl \hspace{0cm} 
&+&\sum_{Q_{s} \sim v} 2i\frac{W([f(Q_{s})]){\rm area}(Q_{s})}{{\rm area}([f(Q_{s})])}
\overline{(f(v'_{s})-f(v'_{s-1}))}\,\partial_{\Lambda}f(Q_{s})\,\bar{\partial}_{\Lambda}f(Q_{s})=0. 
\label{eq_ELeq_fbar}
\end{eqnarray}
Here, we have changed the variables from real $f_{1}$ and $f_{2}$ to complex $f$ and $\bar{f}$ and $v_{0}$ has been replaced by $v$. 

Now, we will show that the discrete holomorphic ($\partial_{\Lambda}f=0$) and anti-holomorphic ($\bar{\partial}_{\Lambda}f=0$) functions satisfy the weighted discrete Euler--Lagrange equation (\ref{eq_ELeq_fbar}). If the function $f$ is discrete holomorphic or discrete anti-holomorphic, then the second sum in (\ref{eq_ELeq_fbar}) vanishes because $\partial_{\Lambda}f(Q_{s})\,\bar{\partial}_{\Lambda}f(Q_{s})=0$. Furthermore, (\ref{eq_e_diamond}) and the identity (\ref{eq_dholo_identity}) yield the relations 
\begin{equation}
e_{\diamondsuit}([f(Q_{s})])=2\rho_{s}\bigl|f(v_{s})-f(v)\bigr|^2=\pm4\,{\rm area}([f(Q_{s})]).
\label{eq_ELeq_fbar=0}
\end{equation}
Therefore, we see that
\begin{equation}
[{\rm EL}]^{\rm disc.}(v) 
=\pm\sum_{Q_{s} \sim v}
\left(\frac{\partial F_{s}\,\,\,\,\,}{\partial \bar{f}(v)}-\frac{\partial F_{s-1}}{\partial \bar{f}(v)}\right)=0, 
\label{eq_ELeq_fbar=0}
\end{equation} 
because the cyclic sum of the differences between the two terms arising from two adjacent sides of the quadrilateral $Q_{s} \sim v$ identically vanishes. Thus, the discrete holomorphic and anti-holomorphic functions satisfy the weighted discrete Euler--Lagrange equation 
(\ref{eq_ELeq_fbar}).

\section{Rhombic lattice}

In this section, we compute the topological quantum number $N$ for discrete power functions defined on 
a rhombic lattice. 
Moreover, we show that the weighted discrete Dirichlet energy 
(\ref{eq_WDEnergy_comp}), the area (\ref{eq_WDArea_comp}), and the discrete Euler--Lagrange equation (\ref{eq_ELeq_fbar}) tend, in a continuous limit, toward their continuous counterparts, namely, the energy (\ref{eq_cont_E}), 
the area (\ref{eq_cont_A}), and the Euler--Lagrange equation (\ref{eq_ELeqf_BG_conti}), respectively. 

Our rhombic lattice $\Lambda$ is shown in Figure \ref{fig:Rhombic_lattice} with lattice spacings $(2a, 2b)$, where $a$ and $b$ are real positive constants. 
\begin{figure}[htbp]
  \begin{center}
   \vspace{1ex}\scalebox{1.0}{\input{Rhombic_lattice_tado.tex}}
   \vspace*{1ex}\hspace*{0cm}\caption{The rhombic lattice $\Lambda$ and 
the even and odd rhombi $Q_{2k,2l}$ and $Q_{2k+1,2l+1}$, respectively} 
\label{fig:Rhombic_lattice}
  \end{center} 
\end{figure}
This $\Lambda$ is a special case of the isoradial graphs in \cite{ChelkSmir11}.  
All rhombi are congruent with each other and their discrete conformal structure is given by $\rho_{Q}=b/a$ for any $Q \in F(\Lambda)$. 
We assume that the left corners of ``even'' rhombi $Q_{2k, 2l}$ are $(2ka, 2lb)$, whereas those of ``odd'' rhombi $Q_{2k+1, 2l+1}$ are $((2k+1)a, (2l+1)b)$, with $(k,l) \in \mathbb{Z}^2$. The set of faces $F(\Lambda)$ is covered by the even and odd rhombi with no overlaps or gaps. 

To derive discrete power functions on $V(\Lambda)$, we begin with the discrete exponential $\mathrm{exp}(\lambda, \cdot; z_0)$ \cite{Mer07, BobeGunt16} 
on $V(\Lambda)$ for the case $z_0=0$ :  
\begin{equation}
\mathrm{exp}(\lambda, z; 0)
=\left(\frac{1+\frac{\lambda}{2}(a+bi)}{1-\frac{\lambda}{2}(a+bi)}\right)^{\frac{x}{2a}+\frac{y}{2b}}
\left(\frac{1+\frac{\lambda}{2}(a-bi)}{1-\frac{\lambda}{2}(a-bi)}\right)^{\frac{x}{2a}-\frac{y}{2b}}
\label{eq_exp}
\end{equation}
with the vertex  $z=x+iy \in V(\Lambda)$ and a complex parameter $\lambda \ne 
\pm 2/(a \pm bi)$. 
This function tends to the $e^{\lambda z}$ in the continuous limit 
$\delta:=\sqrt{a^2+b^2} \to +0$.
The discrete holomorphic power functions $z^{(N)}$ ($N \in \mathbb{Z}_{\ge 0}$) on $V(\Lambda)$ are defined by the expansion  
\begin{equation}
\mathrm{exp}(\lambda, z; 0)=\sum^{\infty}_{N=0}\frac{\lambda^N}{N!}z^{(N)}.
\label{eq_Taylorexp}
\end{equation}
The first few $z^{(N)}$ are $z^{(N)}=z^{N}$ ($N=0, 1, 2$),  
$z^{(3)}=z^{3}+(a^2-b^2)z-\frac{1}{2}(a^2+b^2)\bar{z}$, and 
$z^{(4)}=z^{4}+4(a^2-b^2)z^{2}-2(a^2+b^2)|z|^2$. 
Note that the lattice $\Lambda$ with $a=b=1/{\sqrt{2}}$ and the integer lattice 
$\mathbb{Z}^2$ differ by $\pi/4$ rotation. 
Consequently, the functions $z^{(N)}$ on $V(\Lambda)$ 
agree with Ferrand's discrete power functions on $\mathbb{Z}^2$
\cite{Ferrand} up to multiplication by the factor $e^{-i\frac{N\pi}{4}}$.

If $Q_1$ and $Q_2$ are two quadrilaterals sharing a common edge, 
then we have 
\begin{eqnarray}
\fl &&\sum_{e \in E(\partial Q_1)}\mathrm{area}({\rm St}^{-1}(\triangle{Of(e_{-})f(e_{+})}))
+\sum_{e \in E(\partial Q_2)}\mathrm{area}({\rm St}^{-1}(\triangle{Of(e_{-})f(e_{+})})) \nonumber \\
\fl &=&\sum_{e \in E(\partial (Q_1\bigcup Q_2))}\mathrm{area}({\rm St}^{-1}(\triangle{Of(e_{-})f(e_{+})})). 
\label{eq_Q1UQ2} 
\end{eqnarray}
From (\ref{eq_Sum_area_simpler}), (\ref{eq_cycl_sum_St_inv}), and (\ref{eq_Q1UQ2}), we 
obtain the following formula for finite $\Lambda$
\begin{equation}
{\cal A}^{W}_{\diamondsuit}(f) =\sum_{e \in E(\partial \Lambda)}
\mathrm{area}({\rm St}^{-1}(\triangle{Of(e_{-})f(e_{+})})), 
\label{eq_Sum_area_boundary} 
\end{equation}
which enables us to evaluate the spherical area ${\cal A}^{W}_{\diamondsuit}(f)$ 
by summing along only the boundary $\partial\Lambda$. 

In order to show the relation (\ref{eq_4piN}) for the discrete power functions $z^{(N)}$, 
we define sublattices $\Lambda_{n} \subset \Lambda$ by
\begin{equation}
\Lambda_{n}:=\Bigl\{ z=x+iy \in \Lambda \, \Bigl| \, \Bigl|  
\frac{x}{2a}+\frac{y}{2b}\Bigr| \le n, \, \Bigl|
\frac{x}{2a}-\frac{y}{2b}\Bigr| \le n \Bigr\} 
\label{eq_LambdaL} 
\end{equation}
for integer $n \in \mathbb{Z}_{\ge 0}$. 
The sublattices $\Lambda_{n}$ form an increasing sequence such that $\Lambda_{n} \subset \Lambda_{n+1}$ and 
$\lim_{n \to \infty} \Lambda_{n}=\Lambda$. 
Since the $z^{(N)} \sim z^{N}$ for $z \in V(\partial\Lambda_{n})$ as $n \to \infty$, the relation (\ref{eq_Oz1z2_Sec4}) yields 
\begin{equation}
\mathrm{area}({\rm St}^{-1}(\triangle{Oz^{(N)}(e_{-})z^{(N)}(e_{+})}))
\to 2\mathrm{arg}(e^N_{+}\overline{e^N_{-}})=2N\angle e_{-}Oe_{+}.  
\label{eq_2Nangle} 
\end{equation}
Here $\angle e_{-}Oe_{+}$ denotes the angle between two vectors  
$\overrightarrow{Oe_{-}}$ and $\overrightarrow{Oe_{+}}$. 
Applying (\ref{eq_Sum_area_boundary}) to the dual 
graph $\diamondsuit_{n}=\Lambda^{*}_{n}$, 
we obtain, from (\ref{eq_2Nangle}), the relation (\ref{eq_4piN}) for the 
$z^{(N)}$, i.e,  
\begin{equation}
{\cal A}^{W}_{\diamondsuit}(z^{(N)}) = \lim_{n \to \infty}{\cal A}^{W}_{{\diamondsuit}_{n}}(z^{(N)})=2N\cdot 2\pi=4\pi N, 
\label{eq_to4piN} 
\end{equation}
by means of the formula
\begin{equation}
\sum_{e \in E(\partial \Lambda_{n})} \angle e_{-}Oe_{+}=2\pi.
\label{eq_sum=2pi}
\end{equation}

We consider the continuous limit of our model. 
From now on, we replace the $\Lambda$ 
with its horizontal translation by $z \to z-a$. 
This new rhombic 
lattice makes calculations below simpler in a symmetric way. 
In the continuous limit where the lattice spacings $a$ and $b$ $\to +0$ with a fixed discrete conformal structure $\rho_0:=b/a$, we have for the even rhombi $Q_{2k, 2l}$ 
\begin{eqnarray}
\fl \hspace{0cm} & &\rho_{Q}|f(b_{+})-f(b_{-})|^2+\rho^{-1}_{Q}|f(w_{+})-f(w_{-})|^2 \to 
4ab\left(|f_{x}(2ka,2lb)|^2+|f_{y}(2ka,2lb)|^2\right), \nonumber \\ 
\fl \hspace{.3cm} & &{\rm Im}\left((f(w_{+})-f(w_{-}))\overline{(f(b_{+})-f(b_{-}))}\right) \hspace{.7cm} \to 
4ab\,{\rm Im}\left(f_{y}(2ka,2lb)\bar{f}_{x}(2ka,2lb)\right). 
\label{eq_Climit_even}
\end{eqnarray}
For the odd rhombi $Q_{2k+1, 2l+1}$, we obtain similar results using $((2k+1)a,(2l+1)b)$ instead of $(2ka,2lb)$. The areas of these rhombi are given by ${\rm area}(Q_{2k, 2l})={\rm area}(Q_{2k+1, 2l+1})=2ab$. From these results and the condition (\ref{eq_W(Q)_limt_z}), we see that the weighted discrete Dirichlet energy (\ref{eq_WDEnergy_comp}) and area (\ref{eq_WDArea_comp}) tend to their respective continuous values (\ref{eq_cont_E}) and (\ref{eq_cont_A}):  
\begin{eqnarray}
E^{W}_{\diamondsuit}(f) 
&\to& 2\int_{\mathbb{R}^2}\! d^2x 
\frac{|\partial_{x} f|^2+|\partial_{y} f|^2}{(1+|f|^2)^2}
=\frac{1}{2}\int_{\mathbb{R}^2} d^2x\,\,\partial_{\mu}{\mbox{\boldmath $n$}} \sbt 
\partial^{\mu}{\mbox{\boldmath $n$}}=E, 
\label{eq_climit_E} \\
{\cal{A}}^{W}_{\diamondsuit}(f)
&\to& 4\int_{\mathbb{R}^2} d^2x\,\, 
\frac{{\rm Im}(\partial_{y} f \cdot \partial_{x}\bar{f})}{(1+|f|^2)^2}
=\int_{\mathbb{R}^2} d^2x\,\, {\mbox{\boldmath $n$}}
\sbt (\partial_{x}{\mbox{\boldmath $n$}}\times \partial_{y}{\mbox{\boldmath $n$}})={\cal{A}}.
\label{eq_climit_A}
\end{eqnarray}

We also consider the continuous limit of the weighted discrete Euler--Lagrange equation (\ref{eq_ELeq_fbar}). Applying the Taylor formula for functions of two variables $a$ and $b$ to the signed area (\ref{eq_signed_area}), we have
\begin{equation}
{\rm area}([f(Q_{1})])=2ab{\rm Im}(f_{y}\bar{f}_{x})+2a^2b\partial_{x}{\rm Im}(f_{y}\bar{f}_{x}) 
\label{eq_area_fQ1}
\end{equation}
up to the third order of the lattice spacings, where $Q_{1}$ is the quadrilateral of Figure \ref{fig:Rhombus_Q1-Q4} and $f_{x}:=\partial_{x}f$ and $f_{y}:=\partial_{y}f$. 
\begin{figure}[htbp]
  \begin{center}
   \vspace{1ex}\scalebox{0.8}{\input{Rhombus_Q1-Q4_tado.tex}} \qquad
   \vspace*{-0ex}\hspace*{-3ex}\caption{
Rhombi $Q_{s}$ ($s=1,\cdots, 4$) incident to the vertex $v$ (in the case of black $v$)}
\label{fig:Rhombus_Q1-Q4}
  \end{center} 
\end{figure}
The vertex $v$ is here identified with the complex number $z$. When $f$ is the identity function $f(z)=z$, (\ref{eq_area_fQ1}) reduces to the area of the rhombus $Q_{1}$ (i.e., ${\rm area}(Q_{1})=2ab$). Notably, ${\rm Im}(f_{y}\bar{f}_{x})$ in (\ref{eq_area_fQ1}) is the Jacobian $J$ of the function $f$ at $z$, i.e., $J=\lim_{a,b \to 0}{\rm area}([f(Q_{1})])/{\rm area}(Q_{1})={\rm Im}(f_{y}\bar{f}_{x})$. The discrete derivative (\ref{eq_partial_f_Q}), with $\lambda_{Q}=1/2$, is expanded for $Q_{1}$ and $Q_{2}$ as 
\begin{eqnarray}
\partial_{\Lambda}f(Q_{1})&=&\frac{1}{2}[f_{x}-if_{y}+a(f_{xx}-if_{xy})]=\partial_{z}f+a\partial_{x}\partial_{z}f,
\label{eq_partial_Lambda_Q1} \\
\partial_{\Lambda}f(Q_{2})&=&\frac{1}{2}[f_{x}-if_{y}+b(f_{xy}-if_{yy})]=\partial_{z}f+b\partial_{y}\partial_{z}f, 
\label{eq_partial_Lambda_Q2}
\end{eqnarray}
respectively. The expansions of $\partial_{\Lambda}f(Q_{3})$ and $\partial_{\Lambda}f(Q_{4})$ are obtained from (\ref{eq_partial_Lambda_Q1}) and (\ref{eq_partial_Lambda_Q2}) by the replacements $a \to -a$ and $b \to -b$, respectively. 
For example, the equation (\ref{eq_partial_Lambda_Q1}) is equal to $\partial_{z}f$ evaluated at the center $\diamond_1=z+a=(x+a,y)$ of the quadrilateral $Q_{1}$ up to the first order of the lattice spacings (i.e., $\partial_{z}f+a\partial_{x}\partial_{z}f=\partial_{z}f{\bigr|}_{(x+a, y)}$). Here the symbol $\diamond_{s}$ denotes the center of the rhombus $Q_{s}$ \cite{ChelkSmir11}. Similar observations can also be made in some other expansions.

The first sum on the right-hand side of the weighted discrete Euler--Lagrange equation (\ref{eq_ELeq_fbar}) can be expanded up to the second order of the lattice spacings as 
\begin{eqnarray}
\fl \hspace{2.5cm}& &\sum^{4}_{s=1}\frac{e_{\diamondsuit}([f(Q_{s})])}{\,4{\rm area}([f(Q_{s})])} 
\left(\frac{\partial F_{s}\,\,\,\,\,}{\partial \bar{f}(v)}-\frac{\partial F_{s-1}}{\partial \bar{f}(v)}\right) \nonumber \\
\fl \hspace{2.5cm}&=&-\frac{2iab}{(1+|f|^{2})^{2}L}\Biggl[f_{y}\partial_{x}S-f_{x}\partial_{y}S+\frac{S(f_{x}N-f_{y}M)}{L}\Biggr] 
\label{eq_ELeq_fbar_1st_sum} \\
\fl \hspace{2.5cm}&=&-\frac{2iab}{(1+|f|^{2})^{2}}{\cal D}\Biggl(\frac{S}{L}\Biggr), 
\label{eq_ELeq_fbar_1st_sum_calD}
\end{eqnarray}
where $L$, $M$, $N$, $S$, and ${\cal D}$ are defined by 
\begin{equation}
\fl \hspace{.7cm}L:={\rm Im}(f_{y}\bar{f}_{x}), \,\, M:=\partial_{x}L, \,\, N:=\partial_{y}L, \,\, 
S:=|f_{x}|^{2}+|f_{y}|^{2}, \,\, {\cal D}:=f_{y}\partial_{x}-f_{x}\partial_{y}. 
\label{eq_LMNScalD}
\end{equation}
The differential operator ${\cal D}$ enables us to write the expression (\ref{eq_ELeq_fbar_1st_sum}) in a more concise form (\ref{eq_ELeq_fbar_1st_sum_calD}). In deriving (\ref{eq_ELeq_fbar_1st_sum}), we have used the relation
\begin{equation}
\frac{\partial F_{s}\,\,\,\,\,}{\partial \bar{f}(v)}
=-i\frac{f(v)}{1+|f(v)|^{2}}-i\frac{f(v'_{s})-f(v)}{(1+|f(v)|^{2})^{2}}
+i\frac{4\mu_{s}{\rm Re}(\mu_{s}\bar{f})}{3(1+|f(v)|^{2})^{3}}.  
\label{eq_partialF/partialf}
\end{equation}
This is valid up to the second order of the lattice spacings, with the $\mu_{s}$s denoting  
\begin{equation}
\mu_{1}=-\mu_{3}:=af_{x}+bf_{y}, \quad \mu_{2}=-\mu_{4}:=-af_{x}+bf_{y}. 
\label{eq_mu}
\end{equation}
The relation (\ref{eq_partialF/partialf}) is derived from (\ref{eq_Oz1z2_Sec4}) and (\ref{eq_def_Js}) with the expansion ${\rm tan}^{-1}x=x-x^{3}/3+\cdots$. Note that the last term in (\ref{eq_partialF/partialf}) containing $\mu_{s}$ is finally canceled out each other in deriving (\ref{eq_ELeq_fbar_1st_sum}). 
This cancellation is due to the symmetry of the rhombus $Q_{s}$ under 
the horizontal and vertical reflections across its diagonals. 

Regarding the second sum on the right-hand side of (\ref{eq_ELeq_fbar}), we first expand (\ref{eq_cycl_sum_St_inv}) as 
\begin{equation}
\fl \hspace{.8cm}{\rm area}({\rm St}^{-1}([f(Q_{1})]))=8ab\Biggl[\frac{{\rm Im}(f_{y}\bar{f}_x)
+a{\rm Im}(f_{y}\bar{f}_x)_{x}}{(1+|f(z)|^{2})^{2}}
-\frac{4a{\rm Im}(f_{y}\bar{f}_{x}){\rm Re}(f_{x}\bar{f})}{(1+|f(z)|^{2})^{3}}\Biggr]
\label{eq_St_inv_Q1}
\end{equation}
up to the third order of the lattice spacings. The terms stemming from the cubic term in the expansion ${\rm tan}^{-1}x=x-x^{3}/3+\cdots$ cancel out each other in (\ref{eq_St_inv_Q1}) because of the rhombus symmetry. 
From (\ref{eq_def_W(Q)}), (\ref{eq_area_fQ1}), and (\ref{eq_St_inv_Q1}), we have 
\begin{equation}
W([f(Q_{1})])=\frac{1}{(1+|f(z)|^{2})^{2}}-\frac{4a{\rm Re}(f_{x}\bar{f})}{(1+|f(z)|^{2})^{3}}
\label{eq_WQ1} 
\end{equation}
up to the first order of the lattice spacings. The weight $W([f(Q_{2})])$ is obtained from $W([f(Q_{1})])$ through the replacements $f_{x} \to f_{y}$ and $a \to b$; $W([f(Q_{3})])$ and $W([f(Q_{4})])$ are obtained from $W([f(Q_{1})])$ and $W([f(Q_{2})])$ by the replacements $a \to -a$ and $b \to -b$, respectively. We can easily see that these weights fulfill the condition (\ref{eq_W(Q)_limt_z}). 

The second sum on the right-hand side of (\ref{eq_ELeq_fbar}) is
\begin{eqnarray}
\fl \hspace{.7cm} & &
\sum^{4}_{s=1} 2i\frac{W([f(Q_{s})]){\rm area}(Q_{s})}{{\rm area}([f(Q_{s})])}
\overline{(f(v'_{s})-f(v'_{s-1}))}\,\partial_{\Lambda}f(Q_{s})\,\bar{\partial}_{\Lambda}f(Q_{s}) \nonumber \\
\fl \hspace{.7cm} &=&
\scriptstyle 
\frac{2iab}{(1+|f|^{2})^{2}L}\left[2(|f_{y}|^{2}-|f_{x}|^{2})f_{xy}
+2(f_{x}\bar{f}_{y}f_{xx}-\bar{f}_{x}f_{y}f_{yy})+\overline{(f_{x}N-f_{y}M)}\,\frac{T}{L} \right] 
-\frac{8ab}{(1+|f|^{2})^{3}}\bar{f}T 
\nonumber \\
\normalsize
\fl \hspace{.3cm} &=&
\frac{2iab}{(1+|f|^{2})^{2}}\overline{\cal D}\Biggl(\frac{T}{L}\Biggr)
-\frac{8ab}{(1+|f|^{2})^{3}}\bar{f}T, 
\label{eq_ELeq_fbar_2nd_sum}
\end{eqnarray}
where $T$ and $\overline{\cal D}$ are defined by   
\begin{equation}
T:=(f_{x})^{2}+(f_{y})^{2}, \quad \overline{\cal D}:=\bar{f}_{y}\partial_{x}-\bar{f}_{x}\partial_{y}.  
\label{eq_cal_D}
\end{equation}

Adding (\ref{eq_ELeq_fbar_1st_sum_calD}) and (\ref{eq_ELeq_fbar_2nd_sum}), the weighted discrete Euler--Lagrange equation (\ref{eq_ELeq_fbar}) is expanded as follows: 
\begin{eqnarray}
\fl \hspace{1.7cm}{[{\rm EL}]^{\rm disc.}}(z)
&=&-\frac{2iab}{(1+|f|^{2})^{2}}
\Biggl[{\cal D}\Biggl(\frac{S}{L}\Biggr)-\overline{\cal D}\Biggl(\frac{T}{L}\Biggr)\Biggr]
-\frac{8ab}{(1+|f|^{2})^{3}}\bar{f}T \\ 
&=&\frac{16ab}{(1+|f|^2)^2}\Biggl(\partial_{z}\partial_{\bar{z}}f-\frac{2\bar{f}}{1+|f|^2}
\partial_{z}f \cdot \partial_{\bar{z}}f\Biggr) 
\label{eq_ELeq_fbar_1st+2nd_sum}
\end{eqnarray}
up to the second order of the lattice spacings. In deriving (\ref{eq_ELeq_fbar_1st+2nd_sum}), we have used the relations 
\begin{equation}
{\cal D}S-\overline{\cal D}T-2i(f_{x}M+f_{y}N)=2iL\Delta f, \quad T=4\partial_{z}f \cdot \partial_{\bar{z}}f,
\label{eq_DS-DT}
\end{equation}
where $\Delta=\partial_{\mu}\partial^{\mu}=4\partial_{z}\partial_{\bar{z}}$ is the two-dimensional Laplacian. Finally, from (\ref{eq_ELeq_fbar_1st+2nd_sum}) and (\ref{eq_ELeqf_BG_conti}), we have 
\begin{equation}
{[{\rm EL}]^{\rm disc.}}(z)/(2ab) \to [{\rm EL}]^{\rm cont.}(z),
\label{eq_ELeqclim}
\end{equation}
that is to say, 
the weighted discrete Euler--Lagrange equation (\ref{eq_ELeq_fbar}) 
tends to its continuous form (\ref{eq_ELeqf_BG_conti}) in the continuous limit 
$a,\,b \to0$ for the fixed discrete conformal structure $\rho_0=b/a$. 

\section{Final remarks}

In this paper we present a discrete version of the nonlinear $O(3)$ sigma model by employing the discrete complex analysis on the planar quad-graphs. 
This discrete model has remarkable properties, namely, existence of the topological quantum number and holomorphicity of the solutions; notable characteristics of the continuous counterpart. 
For quad-graphs with orthogonal diagonals, we prove that the discrete (anti-)holomorphic
functions are solutions to the weighted discrete Euler--Lagrange equation.  
We explicitly compute the topological quantum number for the discrete power functions defined 
on a family of rhombic lattices. Moreover, we show that the weighted discrete 
Dirichlet energy, area, and Euler--Lagrange equation tend, in a continuous limit, toward their continuous forms. 

For future research, we hope to extend the model to be defined on quad-graphs with non-orthogonal diagonals. Furthermore, we will study a wider class of functions, for example, discrete meromorphic functions. These issues will be discussed elsewhere. 

\ack
The authors thank T. Sano for helpful discussions about two-dimensional graphs, and also thank F. G{\"u}nther for many valuable comments. \\

\hspace{-0.8cm}{\bf Funding}
\vspace{0.4cm}

\noindent M. K. and Y.T. were supported in part by The Grant for Basic Science Research Projects 
from The Sumitomo Foundation. 
Y. T. was also supported by JSPS KAKENHI Grant Numbers JP25800053 and JP17K05234.

\appendix 
\section{The stereographic projection and the Gauss--Bonnet formula}
Consider a sphere $S^2$ of radius $R$ centered at the origin. The stereographic projection ${\rm St}: S^2 \rightarrow \mathbb{R}^2$ is well-known as 
\begin{equation}
{\rm St}: S^2 \ni (X,Y,Z) \mapsto \frac{R}{R+Z}(X, Y) \in \mathbb{R}^2, 
\label{eq_St} 
\end{equation}
where the reference point is $(0, 0, -R)$. The inverse ${\rm St^{-1}}$ is given by 
\begin{equation}
{\rm St^{-1}}: \mathbb{R}^2 \ni (x, y) \mapsto \frac{R}{R^{2}+x^{2}+y^{2}}(2Rx, 2Ry, R^{2}-x^{2}-y^{2})  \in S^2. 
\label{eq_St_inv} 
\end{equation}
Hereafter, we set $R=1$. The map ${\rm St}$ induces the relationship
\begin{equation}
ds^{2}_{S^2}=\left(\frac{2}{1+x^{2}+y^{2}}\right)^{2}ds^{2}_{\mathbb{R}^2}, 
\label{eq_ds2S=ds2R} 
\end{equation}
where $ds^{2}_{S^2}:=dX^{2}+dY^{2}+dZ^{2}$ and $ds^{2}_{\mathbb{R}^2}:=dx^{2}+dy^{2}$, i.e., the sphere $S^2$ and the plane $\mathbb{R}^2$ are conformally equivalent. 

For a region $F$ in $\mathbb{R}^2$, we have the following formula for the area of ${\rm St}^{-1}(F)$:  
\begin{equation}
\fl \hspace{1cm}{\rm area}({\rm St}^{-1}(F))=\iint_{F} \left|\frac{\partial {\mbox{\boldmath$r$}}}{\partial x}\times \frac{\partial {\mbox{\boldmath$r$}}}{\partial y}\right| dxdy=\iint_{F} \left(\frac{2}{1+x^{2}+y^{2}}\right)^{2}\!\!dxdy, \label{eq_Formula_spherical_area}
\end{equation}
with ${\mbox{\boldmath $r$}}:=(X,Y,Z)$. If $F$ is a triangle $\triangle{Oz_{1}z_{2}}$, we have
\begin{equation}
\fl \hspace{1cm}{\rm area}({\rm St}^{-1}(\triangle{Oz_{1}z_{2}}))=\int^{\theta_{2}}_{\theta_{1}}
\left\{\int^{r(\theta)}_{0}\left(\frac{2}{1+r^2}\right)^{2}rdr\right\}d\theta, 
\label{eq_iint Oz1z2}
\end{equation}
where 
\begin{equation}
\fl \hspace{1cm}r(\theta):=\frac{x_{1}y_{2}-x_{2}y_{1}}{(x_{1}-x_{2})\sin{\theta}+(y_{2}-y_{1})\cos{\theta}},
\quad \theta_{1} \le \theta \le \theta_{2} 
\label{eq_r(theta)}
\end{equation}
is a polar equation of the line segment between the two points $z_{k}:=x_{k}+iy_{k}$, $k=1,\,2$ and $\theta_{k}:={\rm arg}(z_{k})$. A straightforward calculation of (\ref{eq_iint Oz1z2}) yields the formula (\ref{eq_Oz1z2_Sec4}). 
Note that the area (\ref{eq_Oz1z2_Sec4}) is the {\it signed} one, i.e., 
the triangle $Oz_{1}z_{2}$ and ${\rm Im}(z_{2}\bar{z_{1}})=x_{1}y_{2}-x_{2}y_{1}$ 
are counterclockwise or clockwise 
and positive or negative, respectively. The symbol ${\rm tan}^{-1}$ in (\ref{eq_Oz1z2_Sec4}) is the principal value defined such that it lies in the interval $(-\pi/2, \pi/2)$. 

Now, we show that the area (\ref{eq_Oz1z2_Sec4}) can also be derived from the Gauss--Bonnet formula 
\begin{equation}
\int_{M}K\,dA+\int_{\partial M}\kappa_{g}\,ds=\iota_{1}+\iota_{2}+\iota_{3}-\pi, 
\label{eq_GB}
\end{equation}
where $M$ is a spherical triangle, $K$ is the Gaussian curvature of $M$, $dA$ is the surface element of $M$, $\kappa_{g}$ is the geodesic curvature of the boundary $\partial M$, $ds$ is the line element along $\partial M$, and $\iota_{1}$, $\iota_{2}$, and $\iota_{3}$ are the interior angles of $M$. We apply the Gauss--Bonnet formula (\ref{eq_GB}) to the spherical triangle $M={\rm St}^{-1}(\triangle{Oz_{1}z_{2}})$. 
Because the sphere $S^2$ and the plane $\mathbb{R}^2$ are conformally equivalent, angles are preserved under stereographic projection. Hence, we have $\iota_{1}:=\alpha'=\alpha$, $\iota_{2}:=\beta'=\beta$, and $\iota_{3}:=\gamma'=\gamma$ (as shown in Figure \ref{fig:The stereographic projection and the Gauss_Bonnet_formula}). Accordingly, we obtain $\iota_{1}+\iota_{2}+\iota_{3}=\alpha+\beta+\gamma=\pi$ and the right-hand side of (\ref{eq_GB}) vanishes. Because both sides $C_{\beta'}$ and $C_{\gamma'}$ are arcs of the great circles (geodesics), we have$\int_{C_{\beta'}}\kappa_{g}\,ds=\int_{C_{\gamma'}}\kappa_{g}\,ds=0$. As $K=1$, the Gauss--Bonnet formula (\ref{eq_GB}) gives 
\begin{equation}
{\rm area}({\rm St}^{-1}(\triangle{Oz_{1}z_{2}}))=-\int_{C_{\alpha'}}\kappa_{g}\,ds. 
\label{eq_area=-kappa}
\end{equation}

It is known that the geodesic curvature $\kappa_{g}$ of a circle of the latitude of a sphere of radius $a$ is given by $\kappa_{g}=\frac{1}{a}\tan{u}$, where $u$ is the latitude $(-\frac{\pi}{2} < u < \frac{\pi}{2})$. The curvature $\kappa_{g}$ is positive and negative in the upper ($u>0$) and lower ($u<0$) hemispheres of the sphere, respectively. Consider 
the circle $C:=\Pi \cap S^2$, where $\Pi$ is the plane passing through the three points $z_{1}$, $z_{2}$, and the south pole S of the sphere $S^2$. Because the side $C_{\alpha'}$ is an arc of the circle $C$, we have from (\ref{eq_area=-kappa}) 
\begin{eqnarray}
\fl \hspace{1cm} {\rm area}({\rm St}^{-1}(\triangle{Oz_{1}z_{2}}))&=\int_{-C_{\alpha'}}\kappa_{g}\,ds
=\tan{u_{C}}\cdot{\rm length}(C_{\alpha'}), 
\label{eq_tan_length}
\end{eqnarray}
where $u_{C}$ is the latitude of the circle $C$. We change the orientation of $C_{\alpha'}$ to that of $C$.  
\begin{figure}[htbp]
  \begin{center}
   \vspace{1ex}\hspace*{-10ex}\scalebox{1}{\input{stereo_graphic_proj_tado.tex}} 
\vspace*{0ex}\hspace*{-3ex}\caption{The stereographic projection and the Gauss--Bonnet formula} 
\label{fig:The stereographic projection and the Gauss_Bonnet_formula}
  \end{center} 
\end{figure}
The equation of the plane $\Pi$ is given by 
\begin{equation}
\left|
    \begin{array}{llc}
      x & y & z+1 \\
      x_1 & y_1 & 1 \\
      x_2 & y_2 & 1 
    \end{array}
  \right|=0 \label{eq_det_Pi}
\end{equation}
or equivalently $(y_1-y_2)x+(x_2-x_1)y+(x_{1}y_{2}-x_{2}y_{1})(z+1)=0$.  
We define the {\it signed} distance $d_{\Pi}$ {\it from} the plane $\Pi$ {\it to} the origin as 
\begin{equation}
\fl \hspace{.3cm}
d_{\Pi}:=\frac{x_{1}y_{2}-x_{2}y_{1}}
{\sqrt{(y_1-y_2)^2+(x_2-x_1)^2+(x_{1}y_{2}-x_{2}y_{1})^2}}
=\frac{{\rm Im}(z_{2}\bar{z_{1}})}
{\sqrt{|z_{2}-z_{1}|^{2}+[{\rm Im}(z_{2}\bar{z_{1}})]^{2}}}. 
\label{eq_dPi}  
\end{equation} 
Let $r_{C}$ be the radius of the circle $C$. As we have relations 
\begin{equation}
\fl \hspace{3cm} \tan{u_{C}}=\frac{d_{\Pi}}{r_{C}}, \qquad {\rm length}(C_{\alpha'})=r_{C}\cdot2\angle z_{1}Sz_{2}, 
\label{eq_tan=_length=}  
\end{equation}
and 
\begin{equation}
\fl \hspace{1cm}{\rm tan}\angle z_{1}Sz_{2}=\frac{\sin\angle z_{1}Sz_{2}}{\cos\angle z_{1}Sz_{2}}
=\frac{|
\overrightarrow{Sz_{1}} \times \overrightarrow{Sz_{2}}|}
{\overrightarrow{Sz_{1}} \sbt \overrightarrow{Sz_{2}}}
=\frac{\sqrt{|z_{2}-z_{1}|^{2}+[{\rm Im}(z_{2}\bar{z_{1}})]^{2}}}
{1+{\rm Re}(z_{2}\bar{z_{1}})}, 
\label{eq_tan_z1Oz2}  
\end{equation}
the formula (\ref{eq_Oz1z2_Sec4}) is immediately obtained from (\ref{eq_tan_length}) and (\ref{eq_dPi}). 

Berg and L{\"u}scher \cite{BerLu81} derived a formula for the area of the spherical triangle using the Gauss--Bonnet formula. However, there setting is slightly different from ours; in their spherical triangle, all the three sides are arcs of the great circles, but in our spherical triangle, two of the three sides are arcs of the great circles and the remaining one is an arc of the small circle.

\end{document}